\documentclass[format=acmsmall, screen=true, review=false]{acmart}
\usepackage[outercaption]{sidecap}
\usepackage{url}
\usepackage{color}
\usepackage{soul}
\usepackage{xspace}
\usepackage{algorithm}
\usepackage{algpseudocode}
\usepackage{graphicx}
\usepackage{wrapfig}
\usepackage[font={small,it}]{caption}
\usepackage{epstopdf}
\usepackage{comment}
\usepackage{hyphenat}
\sidecaptionvpos{figure}{c}

\usepackage{amsmath}
\usepackage{amssymb}
\usepackage{amsthm}
\usepackage{epsfig}
\usepackage{colortbl}
\usepackage{enumitem}
\usepackage{citesort}
\usepackage{tikz}
\usetikzlibrary{trees,positioning,patterns,fadings}
\usetikzlibrary {shapes}
\usetikzlibrary{decorations,decorations.pathmorphing,decorations.pathreplacing}
\usepackage{algorithm}
\usepackage{algpseudocode}

\usepackage{subfiles}

\usepackage{float}

\usepackage{mathtools}
\DeclarePairedDelimiter{\ceil}{\lceil}{\rceil}

\usepackage{pgfplots}
\pgfplotsset{compat=newest}
\usepgfplotslibrary{colorbrewer}
\usepackage{subcaption}
\newcolumntype{C}[1]{>{\centering}m{#1}}

\newcommand{\br}{\mathbf{r}}

\theoremstyle{definition}

\numberwithin{equation}{theorem}

\begin{document}

%\title{High Performance Evaluation of Helmholtz Potentials using the Multi-Level Fast Multipole Algorithm\tnoteref{t1}}
\title[High Performance Evaluation of Helmholtz Potentials]{High Performance Evaluation of Helmholtz Potentials using the Multi-Level Fast Multipole Algorithm}

%\tnotetext[t1]{This work was supported in part by grants from Riverside Research, Inc. (award \#000000) and the National Science Foundation (award \#1822932). Computational resources were provided by the National Energy Research Scientific Computing Center, (NERSC) and the High Performance Computing Center (HPCC) at Michigan State University.}

\author{Michael P. Lingg}
\affiliation{
	\institution{Michigan State University}
    \department{Computer Science and Engineering}
    \streetaddress{428 S. Shaw Ln.}
    \city{East Lansing}
    \state{MI}
    \postcode{48824}
    \country{USA}
}
\email{linggmic@msu.edu}

\author{Stephen M. Hughey}
\affiliation{
	\institution{Michigan State University}
    \department{Electrical and Computer Engineering}
    \streetaddress{428 S. Shaw Ln.}
    \city{East Lansing}
    \state{MI}
    \postcode{48824}
    \country{USA}
}
\email{hugheyst@msu.edu}

\author{Hasan Metin Aktulga}
\affiliation{
	\institution{Michigan State University}
    \department{Computer Science and Engineering}
    \streetaddress{428 S. Shaw Ln.}
    \city{East Lansing}
    \state{MI}
    \postcode{48824}
    \country{USA}
}
\email{hma@cse.msu.edu}

\author{Balasubramaniam Shanker}
\affiliation{
	\institution{Michigan State University}
    \department{Electrical and Computer Engineering}
    \streetaddress{428 S. Shaw Ln.}
    \city{East Lansing}
    \state{MI}
    \postcode{48824}
    \country{USA}
}
\email{bshanker@msu.edu}

%\cortext[cor2]{Principal corresponding author}

%\address[ece]{Department of Electrical and Computer Engineering, Michigan State University, East Lansing, MI 48824}

%\address[cse]{Department of Computer Science and Engineering, Michigan State University, East Lansing, MI 48824}

\begin{abstract}

Evaluation of pair potentials is critical in a number of areas of physics. The classical $N$-body problem has its root in evaluating the Laplace potential, and has spawned  tree-algorithms, the fast multipole method (FMM), as well as kernel independent approaches. Over the years, FMM for Laplace potential has had a profound impact on a number of disciplines as it has been possible to develop highly scalable parallel algorithm for these potential evaluators. This is in stark contrast to parallel algorithms for the Helmholtz (oscillatory) potentials. The principal bottleneck to scalable parallelism are operations necessary to traverse up, across and down the tree, affecting both computation and communication. In this paper, we describe techniques to overcome bottlenecks and achieve high performance evaluation of the Helmholtz potential for a wide spectrum of geometries. We demonstrate that the resulting implementation has a load balancing effect that significantly reduces the time-to-solution and enhances the scale of problems that can be treated using full wave physics.

% However, in contrast to FMM's well-known Laplace variant, existing FMM implementations for the Helmholtz equation lag far behind in terms of their parallel performance. The principal bottlenecks against large-scale problems are the difficulty of achieving an efficient parallel implementation for tree traversal operations and excessive memory costs, all while ensuring precise control over accuracy. In this paper, we describe techniques to overcome communication bottlenecks and achieve high performance evaluation of the Helmholtz potential for a wide spectrum of geometries. 
%These techniques primarily rely on a cost-effective hybrid scheme to represent the information content in tree nodes, efficient propagation of this data through deep non-uniform FMM trees. 
% We demonstrate that the resulting implementation has a load balancing effect that significantly reduces the time-to-solution and enhances the scale of problems that can be treated using full wave physics.

\end{abstract}

\maketitle
%%%%%%%%%%%%%%%%%%%%%%%%%%%%%%%%%%%%%%%%%%%%%%%%
\section{Introduction}
\label{sec:intro}

Physics described by hyperbolic partial differential equations (PDEs) form the backbone of a wide array of modern technologies. Solutions to PDEs governing electromagnetics and acoustics have enabled technologies that have had, and will continue to have, a broad and profound effect on our daily lives. The common thread for the wide range of phenomena described by the Helmholtz equation is understanding and manipulation of wave physics at multiple length scales. This task is increasingly challenging given the increase in geometric complexity (smaller and more complex features) and wider range of operating frequencies (requiring more precision and detail to achieve optimal performance at all frequency bands). Advances in these technologies, and engineering sophisticated yet robust systems, are intimately tied to a detailed understanding of the underlying wave physics. Today, more often than not, such insight requires simulations via high-fidelity computational tools.

One of the main challenges in developing such tools is computing fields on electrically large objects. Here, electrical length is measured in terms of the largest linear dimension in terms of wavelengths. For several emerging problems, the electrical length can be several thousand wavelengths. Creating a full-wave physics based model for such ``large'' Helmholtz systems represents a major challenge because at its core, a Helmholtz integral equation based solver relies on an iterative sparse solver for which the single most expensive kernel is evaluation of the potential for the corresponding $N$-body problem. The Fast Multiole Method for Helmholtz equations (H-FMM) reduces the $\mathcal{O}(N_{s}^2)$ cost of direct potential evaluation to $\mathcal{O}(N_s \log N_s)$ \cite{Dembart1998} for surface distributions. Here, $N_s$ denotes the number of degrees of freedom. This algorithm bears a strong similarity to that developed for Laplace equations (L-FMM), i.e., for non-oscillatory potentials such as gravitational or electrostatic fields\,\cite{Greengard1987}. The literature on the intricacies of both L-FMM and H-FMM  (and their close cousins, tree-codes) are extensive; see \cite{Shanker2007,Vikram2009a,nishimura2002fast,Appel1985,Barnes1986,Greengard1988}. As is evident from these review papers, applications of these algorithms is extensive and cross-cutting in terms of the number of disciplines that it has benefited. 

Given the wide applicability of FMM, a number of parallel algorithms and parallel implementations have been developed. Those developed for L-FMM have indeed been highly successful in terms of their performance and scalability. Indeed, several Gordon-Bell awards have gone to scalable L-FMM algorithms \cite{Ying2004, hamada2009gordonbell, ishiyama2012gordonbell, rahimian2010gordonbell}. This is in contrast to the development of parallel algorithms for H-FMM, despite sustained efforts \cite{ergul2011parallel,melapudi2011scalable, michiels2013performing, yang2019ternary}. The challenges to developing efficient parallel algorithms for H-FMM arise due to (a) the varying structure and cost of the computational workflow over the underlying tree representation, and (b) accuracy requirements of the target applications. To understand these issues better, we next present the nuances of H-FMM in comparison to L-FMM, summarize the existing literature on parallel algorithms for H-FMM and contributions of this paper that overcome some of the bottlenecks.

%%%%%%%%%%%%%%%%%%%%%%%%%%%%%%%%%%%%%%%%%%%%%%%%

\section{Problem Statement and Background}
\label{sec:background}

Consider a collection of $N_s$ point sources with intensity $u_n\in\mathbb{C}$ located at points $\br_n\in\mathbb{R}^3$, $n=1,\ldots,N_s$. The potential $\Phi(\br)$ due to these sources at some observation point $\br$ is then given by
\begin{equation}
\Phi(\br) = \sum_{n=1}^{N_s} g(\br-\br_n)u_n,\label{eq:helmpot}
\end{equation}
Here, the Green's function $g (\br)$ for the Helmholtz equation is given by
\begin{equation}
g(\br) = \frac{e^{-jk|\br|}}{4\pi|\br|},
\end{equation}
where $k=2\pi/\lambda$ denotes the wavenumber in rad/m and $\lambda$ is the wavelength in meters. Note, the Green's function for the Laplace potential is recovered when $k = 0$. Sums of the form \eqref{eq:helmpot} often arise in the discretization of integral equations in electromagnetics and acoustics \cite{Chew2001}, in which $\Phi(\br)$ must be evaluated at each source point $\br_m,m=1,\ldots,N_s$, implying a cost of $\mathcal{O}(N_s^2)$. The multilevel fast multipole algorithm (MLFMA) \cite{Dembart1995,song1997}, or H-FMM as referred to in this paper, allows approximating these quantities in $\mathcal{O}(N_s\log^2 N_s)$ for surface distributions or $\mathcal{O}(N_s)$ volumetric distributions to arbitrary precision.

\paragraph{Fast Multipole Methods}
Both L-FMM and H-FMM follow the same algorithmic rubric. First, the computational domain is recursively subdivided into cubes (boxes) until a pre-determined box dimension or a maximum number of particles per box is reached, and each particle is mapped to the box in which it resides. The hierarchical structure resulting from this recursive subdivision procedure can be represented as a tree %(see Fig. \ref{fig:tree}), 
specifically an \emph{octree} in which each subdivision results in the creation of (up to) eight boxes of half the diameter in 3D.

%\shankernote{I do not think we need this at this point -- detracts from contribution}
%\stevenote{The basic unit of computation in the MLFMA is the plane wave spectrum (PWS), which may w.l.o.g. be thought of as a two-dimensional array consisting of samples of a function on the unit sphere, parametrized by the variables $(\theta,\phi)\in\left[ 0,\pi \right] \times \left[ 0,2\pi \right)$. The PWS for a source box is called a multipole expansion; the PWS for a target box is called a local expansion.} 

\begin{figure}
	\centering
	%\vspace{-0.2in}
	\includegraphics[width=0.5\linewidth]{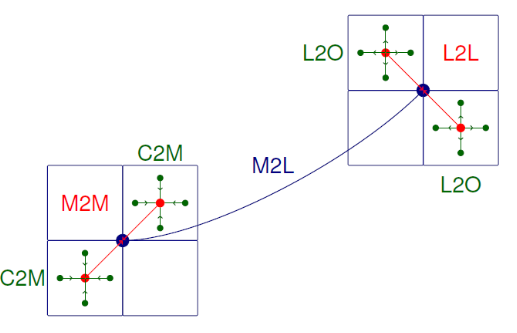}
    %\vspace{-0.1in}
	\caption{Top-down view of the lowest two levels of an FMM tree, illustrating the key computational kernels. Particles (green dots) are mapped onto leaf nodes (small squares) in an octree partitioning.}
	%\vspace{-0.in}
	\label{fig:fmm}
\end{figure}

Assuming an octree with $L$ levels, let the root node reside at level $l=1$, and the leaf boxes reside at level $l=L$. At all levels, boxes are classified as being either in the near- or far-field of each other if they are sufficiently close. Two boxes are in each other's near-field if their domains share a face, edge or node. At any given level, two boxes are designated to be in each other's far-field if (a) they are not in the near field of each other, and (b) their parents are in each other's near-field. This permits a hierarchical partitioning of the computational domain in terms of near- and far-field interactions. Near-field calculations take place \emph{only} at the lowest level; at all other levels, interactions between boxes is performed via far-field through a five stage process shown in Figure\,\ref{fig:fmm} and listed below (note, specific mathematical details are omitted in this discussion, and can be found in \cite{Greengard1998,Wandzuraz1993} and elsewhere \cite{Vikram2009a,nishimura2002fast}):
\begin{enumerate}
	\item Compute charge to multipole information for each leaf node based on the particles it encloses (C2M), 
	\item compute the multiple expansions for each node in the tree by interpolating from the multipole information of all of its children (M2M),
	\item calculate interactions between far-field pairs by translating multipole expansions of sources to the observers' locations (M2L),
	\item starting at the highest level nodes and going all the way down to the leaves, distribute (anterpolate) multipole expansions aggregated at non-leaf observer boxes as a result of M2L (these expansions are then referred to as \emph{local expansions}) to their children (L2L),
	\item convert the resulting local expansions at each leaf box to particles enclosed therein (L2O).
\end{enumerate}

\paragraph{Laplace vs. Helmholtz FMM and Ramifications on Parallelization} As alluded to earlier, highly efficient parallel implemetations of L-FMM exist \cite{ishiyama2012gordonbell,rahimian2010gordonbell} in contrast to those for H-FMM \cite{ergul2013fast,hughey2019parallel,michiels2013weak,yang2019ternary}. To understand why, one needs to examine the underlying mathematics.
\begin{itemize}
  \item In L-FMM, irrespective of the tree level, the amount of multipole data needed to be stored at each node in the tree is identical because the magnitude of a Laplace potential monotonically decreases with distance. 
  \item Due to the constant information content in L-FMM nodes, the per-level cost of far-field interactions falls off exponentially, and the bulk (80-90\% or more) of the work in L-FMM is associated with the leaf nodes. Thus, existing L-FMM efforts have focused on optimizing the computation and masking communication costs primarily at the leaf level\,\cite{ying2004kernel,sundar2008bottom,agullo2014task}.
  
  \item In contrast, in H-FMM the amount of multipole data associated with a node depends on the level due to the oscillatory nature of the potential. More precisely, the amount of multipole data in a parent must be at least four times larger than that of its children to ensure that the desired level of accuracy can be obtained. This fundamental difference leads to major scalability and memory bottlenecks in H-FMM implementations \cite{hughey2018parallel,melapudi2011scalable}.
  %\item Additionally, as one ascends an octree ($l\rightarrow l+1$), at each level the number of nodes decreases by (roughly) a factor of {\em four} for surface geometries, and {\em eight} for volumetric systems.
  
  \item Due to the quadrupling of the information content as one ascends an H-FMM tree, computational and memory costs in H-FMM stay constant for each tree level for surface geometries  and halves for volumetric problems. For this reason, advances in parallel L-FMM do not translate readily to parallel H-FMM despite significant and sustained research \cite{ergul2011solutions,ergul2013accurate,ergul2013fast,Vikram2011,waltz2007massively,taboada2013mlfma}. The commonly-used local essential tree (LET) \cite{salmon1991parallel} paradigm from the L-FMM literature fails for H-FMM because the size of the "ghost" region representing sources residing on other processes rapidly exceeds available memory.
\end{itemize}

Addressing these performance and scaling challenges, while maintaining high accuracies in potential evaluations, constitutes the main motivation for the present work.

% FMM for the Laplace equation (L-FMM) has been explored in depth, to a large extent driven by gravitational physics \cite{Warren1993} \cite{Greengard1997}, electrostatics \cite{}, and others \cite{Vikram2009}. Indeed, a number of highly efficient parallel implementations exist \cite{}\shankernote{Need some citations}. While significant efforts have been expended in parallelizing Helmholtz FMM (H-FMM) \cite{}, the state of art is nowhere close to that of L-FMM. The root of this problem is the nature of the H-FMM algorithm, and in turn, the physics arising from these equations. The potentials associated with the Laplace equation are non-oscillatory and inversely proportional to the distance between two points. As a result, the number of multipoles chosen needs only to accurately reconstruct the \textit{magnitude}, and is fixed for \emph{all} nodes in a tree. In stark contrast, for Helmholtz systems, the number of multipoles {\em must} be chosen so as to accurately represent \textit{both phase and magnitude} due to the oscillatory nature of the potential. Therefore, the information content in an H-FMM node is proportional to the square of the node's electrical/acoustic size. As the electrical size at level $l+1$ is double that of $l$, the information content at $l+1$ is four times that at $l$. \shankernote{check notation -- if not fix and delete note}. 

\paragraph{Related Work} 
%While efficient parallelization techniques for H-FMM have been well-studied \cite{velamparambil2000parallelization,ergul2008hierarchical,michiels2013weak,hughey2019parallel}, scalability results for these methods have certainly not approached those of parallel L-FMM codes, a direct consequence of the H-FMM algorithm.  
The aforementioned work profile of the H-FMM octree  suggests that any efficient parallelization must strike a balance between distributing the many lightweight boxes at lower levels and distributing the work of the few heavyweight boxes at higher levels across processes. Furthermore, the mathematics used to effect each stage of the process dictate the intricacies of the algorithms developed for parallelization. The existing algorithms address these different scenarios with different trade-offs. For the purposes of the ensuing discussion, we note that multipole and local expansions may be viewed as two-dimensional arrays of sampled function values.

At scale, the multipole and local expansions of octree boxes at the uppermost levels of the tree must be distributed across processes to achieve load balance \cite{velamparambil2000parallelization}. To reduce the costs of communication in M2M and L2L for these distributed nodes, several authors have employed \emph{local} interpolation techniques \cite{chew2001fast,ergul2008hierarchical,michiels2013weak}, in which only a small "halo" of nearby samples are required to calculate each new sample. However, despite slightly reducing the H-FMM's asymptotic complexity from $\mathcal{O}(N_s\log^2N_s)$ to $\mathcal{O}(N_s\log N_s)$ \cite{cecka2013fourier}, this approach increases memory and computational costs stemming from the need to oversample multipole and local expansions, in addition to reducing the numerical accuracy of the H-FMM. Alternatively, we propose the use of a parallel version of the \emph{global} interpolation method, which has typically remained restricted to the serial M2M/L2L operations at lower levels of the tree. Though the global algorithm obviously has higher communication costs, it does not introduce additional numerical errors, and it facilitates optimal (minimum) sampling rates for multipole/local expansions\,\cite{hughey2019parallel,lingg2018optimization}.

Local interpolation based hierarchical partitioning (HiP) approach distributes expansions hierarchically at the uppermost levels in block columns, or strips~\cite{ergul2008hierarchical}. However, the M2M and L2L communication costs scale as $\mathcal{O}(\sqrt{N_s})$ per process, hampering the scalability of the H-FMM evaluation~\cite{ergul2013fast}. The blockwise HiP (B-HiP) strategy \cite{michiels2011towards,michiels2013weak} alleviates this bottleneck by distributing expansions in blocks, whose much lower surface-to-volume ratio results in $\mathcal{O}(1)$ communication costs per process, hence improving scalability \cite{michiels2015full}. In both methods, M2L operations are carried out in parallel by collecting on each process samples of the remote multipole expansions required to compute the local expansions it owns. It should also be noted that the increased sampling required with local interpolations hampers the scalability of the M2L phase, as collecting remote multipole expansions requires a significantly higher communication bandwidth compared to a global scheme with optimal sampling rates.

Building on the HiP approach, Yang et al. \cite{yang2019ternary} transition from hierarchical partitioning to plane-wave partitioning (PWP) \cite{velamparambil2000parallelization} for the highest levels of the tree, using a binary tree decomposition of the MPI communicator to flexibly load balance the computation. The PWP approach achieves zero communication overhead for M2L operations by distributing expansions at the uppermost levels of the tree by assigning each process a fixed window of samples for \emph{all} expansions at a given level. However, the transition from HiP to PWP requires expensive communications in M2M and L2L phases to rearrange the expansions, though this cost may be justified by recognizing that each node interacts with at most 8 other nodes to perform the M2M/L2L operations, while the maximum number of nodes for M2L operations is 189 (with a volumetric problem).

As previously stated, local interpolation methods are convenient for parallelization but result in an H-FMM that is \textbf{not} strictly error-controllable. The principal challenge to a scalable H-FMM with error control is the communication cost of distributed global (exact) interpolation. In \cite{melapudi2011scalable}, Melapudi\,et al. describe an error-controllable H-FMM based on global interpolation using a bottom-up partitioning which gives great flexibility for load balanced partitioning of the tree. Scaling of this implementation is hampered by the coarse-grained parallelization of the M2L phase and redundant M2M/L2L calculations associated with high-level tree nodes shared by multiple processes.

\paragraph{Contribution}  
%The objective of this work is to develop an efficient parallel MLFMA implementation with controllable accuracies for electrically large domains that comprise non-uniform discretizations. While it is known that a particular class of tree traversal operators can ensure controllable accuracy \cite{Melapudi2011,Sarvas2003}, their use brings a set of challenges, especially for deep H-FMM trees. As elucidated in Sect.\,\ref{sec:algo_base}, the critical bottlenecks that arise are (a) memory requirements, and (b) load balancing deep non-uniform H-FMM trees. In this paper, we utilize a hybrid tree traversal scheme (see Section\,\ref{sec:hybrid}) which significantly reduces the overall memory requirements and we present an efficient parallelization of this hybrid scheme. Then in Section\,\ref{sec:loadbalancing}, we present an empirical load balancing tehcnique to ensure scalable computations with electrically large objects. As shown through extensive numerical tests in Sect.\,\ref{sec:performance}, our proposed techniques achieve significant speedup over existing approaches, all with controllable accuracy.
In this paper, we build upon our earlier efforts \cite{melapudi2011scalable,hughey2019parallel} toward a scalable, error-controllable H-FMM based on global interpolation. %This line of research implies that  our traversal up and down the tree is via global interpolation for reasons elucidated in \cite{hughey2019parallel}.
We address several challenges regarding parallelization and communication, and we demonstrate an efficient and scalable method for evaluation of the Helmholtz potential. In particular, our contributions can be summarized as follows:
\begin{enumerate}
  \item Development of a fine-grain parallel algorithm with bottom-up partitioning that enables scalable evaluation of deep uniform MLFMA trees,
  \item maintain the high level of controllable accuracy shown in previous global interpolation implementations,
  \item a detailed analytical model to characterize the complexity and memory use of the parallel algorithm for far-field interactions,
  \item and, demonstration of the overall algorithm performance and validation of this performance against our analytical model for different test scenarios. 
  %and accuracy of the analytical model under a number of different test scenarios.
\end{enumerate}

\section{Parallel Algorithms and Implementations}
\label{sec:algo_base}

%Since the information content of a tree node in MLFMA increases by factor of 4 while going up the tree, computational and memory loads at \emph{each} MLFMA tree level is significant, and need to be distributed in a balanced way for a scalable evaluation. This is especially crucial for high levels of the tree where there gradually are fewer but larger tree nodes as one ascends the tree. In comparison, high level tree nodes are not a concern in L-FMM, as the total information content across all nodes in a level decreases exponentially while moving up an L-FMM tree. Due to the strict data dependencies in MLFMA (between children and parents in M2M/L2L, between source nodes and observers during M2L), assigning high-level tree nodes to separate processes is not an efficient parallel execution option. To fully leverage the advantages of global interpolation, a fine-grained parallel approach is necessary. For this purpose, we adopt the bottom-up partitioning approach of Melapudi et al.\,\cite{melapudi2011} due to the flexibility it provides in partitioning the MLFMA tree among processes for non-uniform geometries.

In what follows, we describe details of each stage of the algorithm. For completeness, we replicate some of the concepts introduced in \cite{melapudi2011scalable,hughey2019parallel}, before delving into details of our specific contributions.

% As in our earlier work \cite{melapudi2011scalable,hughey2019parallel}, our interest is in creating a scheme whose accuracy is controllable. To this end, our numerics will use global interpolation/anterpolation for M2L/L2L operations. To fully leverage the advantages of global interpolation, a fine-grained parallel approach is necessary. For this purpose, we adopt the bottom-up partitioning approach due to the flexibility it provides in partitioning the MLFMA tree among processes for non-uniform geometries. In what follows, we introduce these concepts before delving into details of our contributions. 

\subsection{Tree construction and setup}
Let $N_p$ denote the number of processes used in the computation. We initially distribute the $N_s$ particles evenly across all processes and determine the diameter $D_0$ of the cube bounding the entire computational domain. Given the finest box diameter $D(L)$, the number of levels in the tree is calculated as the smallest integer $L$ such that $L\geq \log_2(D_0/D(L))+1$. Once the number of levels and therefore the position of the leaf nodes are known, every particle is assigned a key based on the Morton-Z order traversal of the MLFMA tree \cite{Warren1993}. A parallel bucket sort on the Morton keys is then used to roughly equally distribute particles across processes at the granularity of leaves. This is done by selecting $N_p-1$ Morton keys, or ``splitters'', which chop the Morton Z-curve into $N_p$ contiguous segments. Whole leaves are uniquely assigned to processes using these splitters. Given a contiguous segment of leaf nodes, each process determines all ancestor keys of its leaves up to the root. The leaf through ancestor nodes are used to construct the \emph{local subtree}. A simple method of storing the local subtree is as a post-order traversal array. To quickly access a random node, we use an indexer into this local subtree array.

\paragraph{Plural Nodes}

\begin{figure}
\centering
\includegraphics[width=0.9\linewidth]{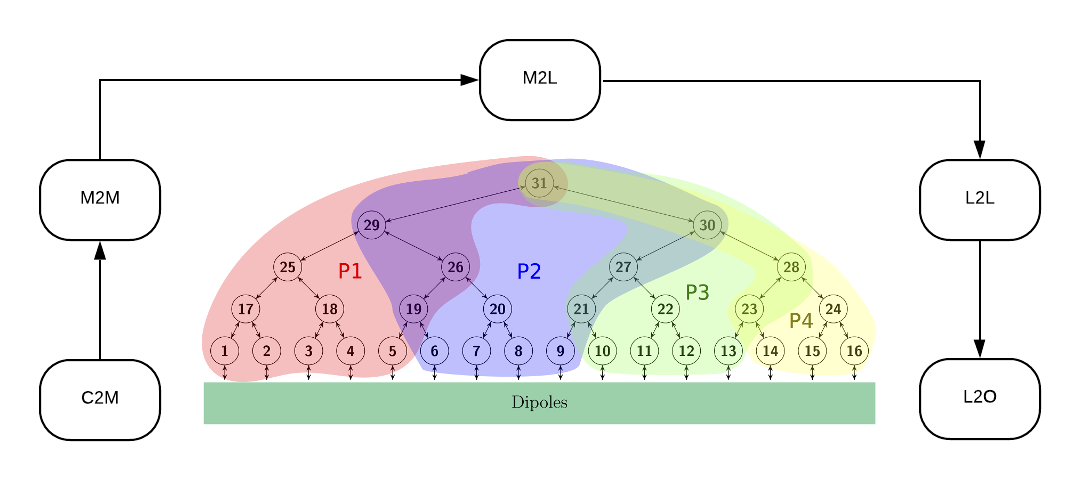}
\caption{Parallel Farfield MLFMA.}
\label{fig:tree}
\end{figure}

Despite the non-overlapping partitioning of leaf nodes, overlaps among different processes at the higher level nodes are inevitable (and in fact, are desirable) as illustrated in Fig.\,\ref{fig:tree}. Details and associated proofs on such partitioning can be found in \cite{melapudi2011scalable}. We call such nodes shared by multiple processes as \emph{plural nodes}. While there are no limitations to the number of processes that can share a plural node, we designate a particular process, i.e., the right-most process sharing the plural node in the MLFMA tree, as its \emph{resident process}. We refer to the resident process' copy of a plural node as a \emph{shared node} and all other copies of this node residing on other processes as \emph{duplicate nodes}. We call the set of processes that own these duplicate nodes as \emph{users} of the shared node, denoted by $U(s)$, where $s$ is the shared node.

One notable advantage provided by plural nodes is that storage of the node is split between multiple processes. In this case, the indexer additionally stores which \emph{slice} of a tree node the current process actually stores in its local subtree array. As the information content for a node is not available to any single process in its entirety, a fine grain parallelization is necessary to perform computations associated with plural nodes. We note that a process can have at most two plural nodes per level in its local tree (essentially one to the ``left'', and another to the ``right'').

\subsection{Parallel Evaluation}

%Figure\,\ref{fig:treePart} gives an example of the bottom-up partitioning where leaf nodes correspond to non-empty boxes, \emph{i.e.}, those containing particles. As a simple heuristic to balance the load, a distinct set of contiguous leaf nodes is assigned to each process such that particles are partitioned as evenly as possible among processes (to the extent allowed by the distribution of particles into leaf nodes). During MLFMA computations, each process is responsible for all leaf level nodes assigned to it, as well as the ancestors of these nodes. 

%Of the five different phases in the MLFMA algorithm, C2M and L2O phases occur at the leaf level where each process performs the necessary computations on particles they are responsible for in an embarrassingly parallel manner. Given the high leaf nodes to process ratio and the fact that no plural nodes can be present at the leaf level, C2M and L2O phases can easily be parallelized with high efficiencies. Therefore, our discussion will focus on M2M, M2L and L2L phases -- together, they account for the vast majority of the total computation time. \textcolor{red}{Are we deviating from our algorithm for C2M and L2O --- }. 

\subsubsection{C2M}
\label{sec:m2m}
C2M is unchanged from our previous works \cite{melapudi2011scalable,hughey2019parallel}.
As each process is assigned a contiguous segment of whole leaf nodes, each process handles the C2M phase for its assigned leaf nodes in parallel independently.

\subsubsection{M2M}
\label{sec:m2m}

In a nutshell, M2M creates multipole expansions of non-leaf boxes from the multipole expansions of their children. This first requires multipole expansions of all children to be \emph{interpolated} to the size of the parent box. Next, each interpolated child box is \emph{shifted} from the center of the child box to the center of the parent box. Finally, multipole contributions of every shifted child box are \emph{aggregated} to form the multipole expansion for the parent box.

M2M computations start at the leaf level and proceed upwards in the tree following a post-order traversal of the local tree. Our parallelization of \emph{M2M} depends on the level of the node and is described in  Alg.\,\ref{alg:m2m}. The approach is as follows: i) Non-plural tree nodes (that are typically found at lower levels of the tree) are handled by their owner processes in parallel independently, ii) for plural nodes without any plural children, \emph{interpolation} and \emph{shift} steps for child nodes are performed sequentially, and the aggregation step is performed as a reduce-scatter operation among processes sharing the plural node, iii) plural nodes with plural children (which typically are located at the higher levels of the tree and incur significant computational and storage costs) are processed using a fine-grained parallel algorithm that we discuss in more detail below. 

\begin{algorithm}
 \caption{Multipole-to-multipole interpolation}
  \begin{algorithmic}[1]
  \Require $p.center$ coordinates of the parent box center
  \Ensure $pmp$ is parent's multipole representation
  
  \For{each box $p$ in post-order traversal}
   \For{each child box $c$}
    \If{$c$ is plural}
     \State $mp[c]\gets$ parallel\_interpolation($c$)
    \Else
     \State $mp[c]\gets$ serial\_interpolation($c$)
    \EndIf
   \State $smp[c]\gets$ shift($mp$, $p.center$)
  \EndFor
  \If{$p$ is plural}
    \State reduce\_scatter($smp$, $users(p)$)
    \Else
    \For{each child box $c$}
      \State aggregate($pmp$,$smp[c]$)
    \EndFor
    \EndIf
 \EndFor
 \end{algorithmic}
 \label{alg:m2m}
\end{algorithm}

\paragraph{Fine-grained Parallel Interpolation}
For plural nodes that necessitate fine-grained parallelization of M2M, the multipole data of the child nodes needed for FFTs are themselves split among multiple processes as indicated in Alg.\,\ref{alg:m2m}. Prior to elucidating our parallel algorithm, we note that our M2M implementation uses a Fast Fourier Transform (FFT)-based interpolation over the uniformly spaced multipole expansions of the child nodes \cite{Sarvas2003}. FFT-based interpolation on the Fourier sphere requires equispaced samples along $\theta$ (vertical) and $\phi$ (horizontal) directions. Due to the inter-dependencies of the FFT algorithm, there is no way to partition the data so as to avoid communication. 

\begin{figure}
\centering
\includegraphics[width=\linewidth]{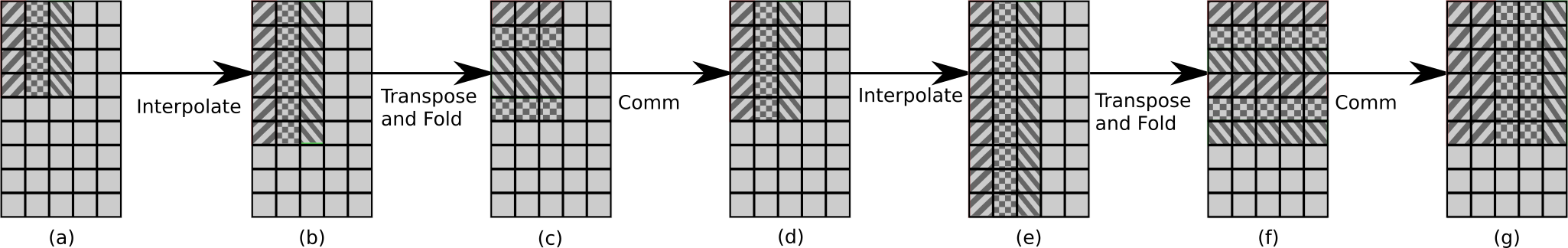}
\caption{Graphical illustration of the transposition and folding operations during fine grained parallel interpolation of multipole data of child node $c$ to parent node $p$ for $N_\theta=3,N_\phi=4$ (for $c$) and $M_\theta=5,M_\phi=6$ (for $p$) using 3 processes each of which owns a column of the initial multipole data as indicated by the hashmarks. Multipole data from (a) is interpolated along the $\phi$ dimension locally, leading to the multipole expansions in (b). The folding operation acting on the interpolated data is shown by the repositioning of the data as in (c). The hash marks show how the folded data is stored on the wrong processes, and must be communicated to the correct process as shown in (d). With the entire multipole data columns in the $\theta$ direction being available on each process, another set of interpolations are performed locally (e), which is then transposed and folded to yield the final multipole expansions (f). A final communication step is needed to send each $\theta$ vector to their owner processes (g) which can then be shifted to the center of the parent box and added to the parent's multipole expansions in accordance with the spherical symmetry condition.}
\label{fig:parallel_interpolation}
\end{figure}

Our approach is as follows: First, each process is assigned a (roughly) equal number of contiguous columns of multipole data (which correspond to groups of samples along the $\phi$ direction). 
%Then each process is assigned one extra column of $\phi$ data, to assign all remaining vectors. So the first $n$ processes will each own $\ceil{\frac{N_\theta}{P}}$ columns of $\phi$ data where $n=\bmod \frac{N_\theta}{P}$ and the remaining processes will each own $\floor{\frac{N_\theta}{P}}$ columns of $\phi$ data. 
The operation begins with a set of independent FFTs along these columns for interpolation in the $\phi$ direction, performed the same as the basic Sarvas approach. Then, the interpolated columns are shifted into rows (see Fig.\,\ref{fig:parallel_interpolation}), transposing and folding the samples in the $\theta$ direction into individual columns. The next step with serial processing would be FFT interpolation in the $\theta$ direction, but this data is split between processes sharing the plural node. 
%The transpose and fold results in each process storing data in columns outside those assigned to the process, and some data in each process' columns being owned by other processes. 
Therefore, each process is communicated the  $\theta$ samples they need to complete their assigned columns using an \verb|MPI_Alltoallv| collective call. Now that each process is storing full columns of $\theta$ samples, these multipole data can be interpolated. The fully-interpolated multipole data is transposed and folded back to its original form ($\phi$ samples along columns, $\theta$ samples along rows). The data are again communicated back to the processes that own the corresponding multipole samples via another \verb|MPI_Alltoallv|. These major steps are illustrated in  Figure\,\ref{fig:parallel_interpolation}.

\paragraph{Shifting of Interpolated Data}
Shifting of multipole data is simply the translation of the interpolated samples from the child node's center to the parent node's center. Translation of each multipole data is independent of others and trivially parallelizable even in the case of fine-grained parallel M2M.

\paragraph{Aggregation}
Aggregation requires adding all corresponding samples from each interpolated and shifted child node together to form the multipole expansion of a parent node (step (g) in Fig.\,\ref{fig:parallel_interpolation}). When children are owned by separate processes (as is the case for plural nodes), reduction communications are required. Note that in fine-grained parallel M2M for a plural node, each process owns only a portion of the parent node's multipole data. A reduce/scatter operation (for instance using \verb|MPI_Reduce_scatter|) could perform both the aggregation and distribution of the appropriate portions of the aggregated multipole data to the processes sharing a plural node. One complication here is that the reduce/scatter operation would require memory to be allocated to the full-size of the parent node by each user process through padding the parts not owned by a process with 0s. Clearly, this would lead to significant memory and computational overheads, especially at the highest levels of the H-FMM tree (due to the large sizes of the plural nodes there). Therefore, we opt for a custom point-to-point aggregation scheme where the interpolated and shifted multipole samples from child nodes are communicated directly (via \verb|MPI_Send| and \verb|MPI_Recv|s) to the process that owns the corresponding samples of the parent node. If the source and destination are the same process, obviously no communication is performed. Each process sharing the parent node then adds up the corresponding multipole samples it receives from each child node, local or communicated. This method reduces both the amount of temporary memory necessary for aggregation and the overall communication volume.

\begin{figure}
	\centering
	\includegraphics[width=0.75\linewidth]{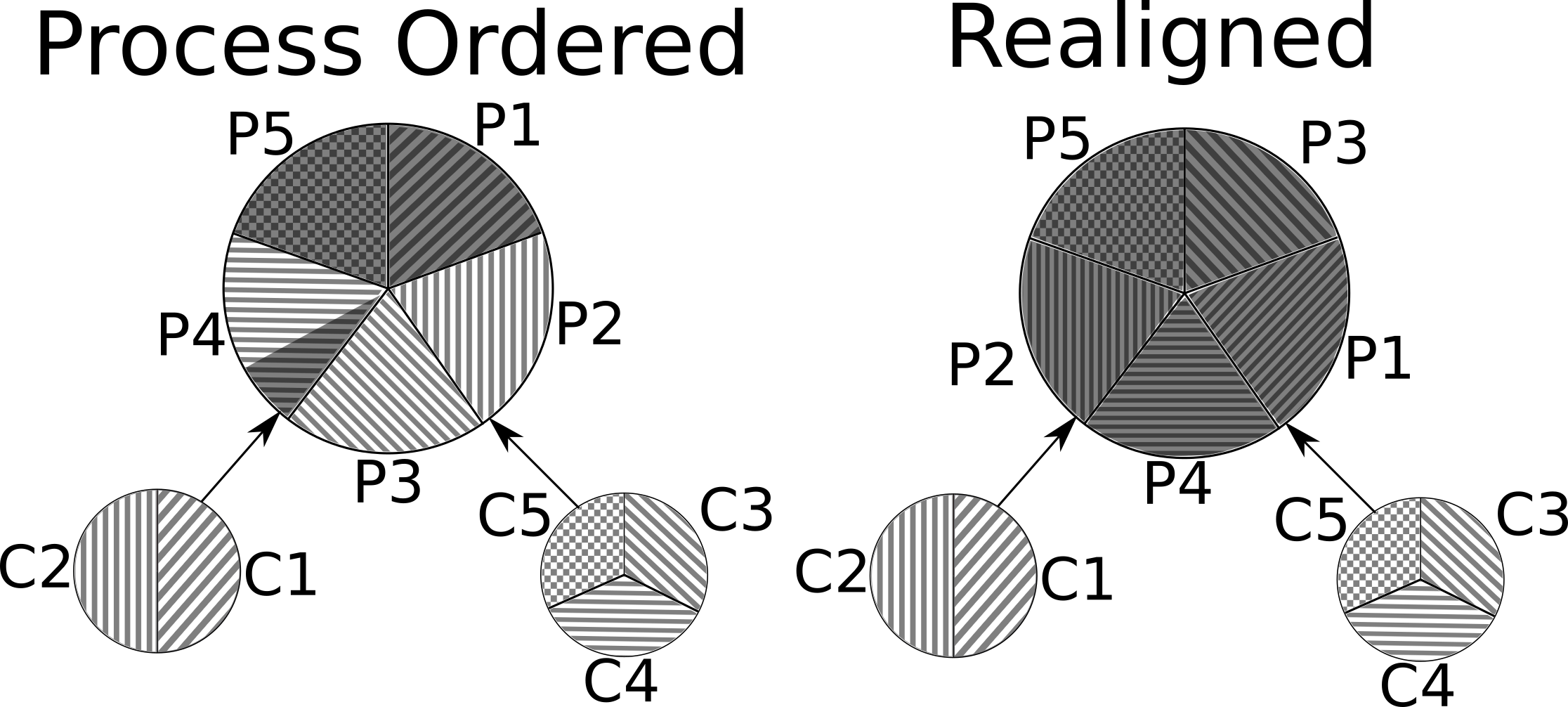}
	\caption{This image shows how multipole samples, ordered clockwise starting at 12 o'clock, are assigned to processes owning the children nodes (C1-C5) overlap with processes owning the parent node (P1-P5) when assigned in process rank order on the left, and with our realignment scheme on the right. The darker portions on the parent samples show the regions where the parent node data overlap with the child node data, and are essentially local data that do not require communication.}
	\label{fig:process_alignment}
\end{figure}

\paragraph{Process Alignment}
In fine-grained parallel M2M, performance impact of how the multipole data of child and parent plural nodes are mapped to processes sharing those nodes also needs to be considered. From the description of our custom point-to-point aggregation scheme above, it is evident that increasing the overlap of the multipole sample regions owned by a process in the child and parent nodes is critical for reducing the communication volume. 
%During M2M of plural nodes, interpolated nodes are aggregated together, then distributed among the processes that share the node. If some of the multipole samples of the parent node a process owns corresponds with the samples the process owns in the child node, the corresponding interpolated and shifted samples from child node are already local to this process. So this data does not need to be communicated to another process to aggregate the data into the parent node. 
As an extreme example, if a process owns no samples in the parent node that correspond with any of the samples it  owns in the child node, all its interpolated and shifted child node samples would have to be communicated to another process for aggregation. In fact, this extreme example is not uncommon when multipole data of a node is simply partitioned into blocks and mapped to user processes according to their process ranks. This situation is illustrated on the left side of Fig.\,\ref{fig:process_alignment}; some processes own samples of a parent node that has no overlap with the samples they own in the child node. Specifically, while process 1 owns overlapping samples in the child and parent nodes, process 2 and 3 own no overlapping samples. 

As a heuristic to minimize the communication volume, we order processes within a parent node such that the parent node samples are assigned by following the priorities below to ensure maximal overlap with their child node samples:
\begin{enumerate}
  \item Index of the lowest sample they own in the child nodes (lower comes first),
  \item number of samples they own in the child nodes (fewer comes first),
  \item process rank.
\end{enumerate}

In the example given in Fig.\,\ref{fig:process_alignment}, both process 1 and process 3 own samples with index 0 in the child nodes, but process 3 has a smaller number of samples so it is assigned the first portion of the parent node samples with process 1 being assigned the second portion. Following processes 3 and 1, process 4 owns the multipole sample with the lowest index in the child nodes, followed by process 2, and then process 5. As such, remaining portions of the parent node samples are assigned in this order. As can be seen in the figure, all samples each process owns in the parent node fully overlap with samples that they own in the child nodes, despite the non-uniform layout. With the proposed process alignment scheme, process 3 will still need to communicate some samples to process 1 for aggregation, but over half of the child samples interpolated by process 3 remains local. Note that with the straight-forward ordering of processes by their ranks, the entire child data interpolated by process 3 would have to be  communicated to processes 1 and 2. In the new scheme, all processes use interpolated samples from their part of a child node without having to communicate. While this scheme would work best with a perfectly balanced tree, this approach will still be effective in reducing the communication volume during aggregation with any tree structure.

\subsubsection{M2L}
\label{sec:m2l}

The M2M step builds the multipole expansions of all tree nodes owned/shared by a process, starting from the leaves all the way up to the root (or the highest level of computation). %Multipole to local (M2L) translation stage is where all far-field interactions actually take place. In other words, M2L translates and applies the multipole data from \emph{source} nodes to \emph{observer} nodes, effectively creating the potential observed locally in each node, i.e., the \emph{local expansions}.
During M2L each observer node loops through all source nodes in its far-field and translates the source multipole data to its locale, aggregating the effects from all its far-field interactions in the process. When the source-observer node pair is on the same process, this interaction is handled purely locally. However, when the source node data is on another process (or a set of processes), one needs a load balanced algorithm that is communication efficient. 

\begin{figure}
	\centering
	\includegraphics[width=0.35\linewidth]{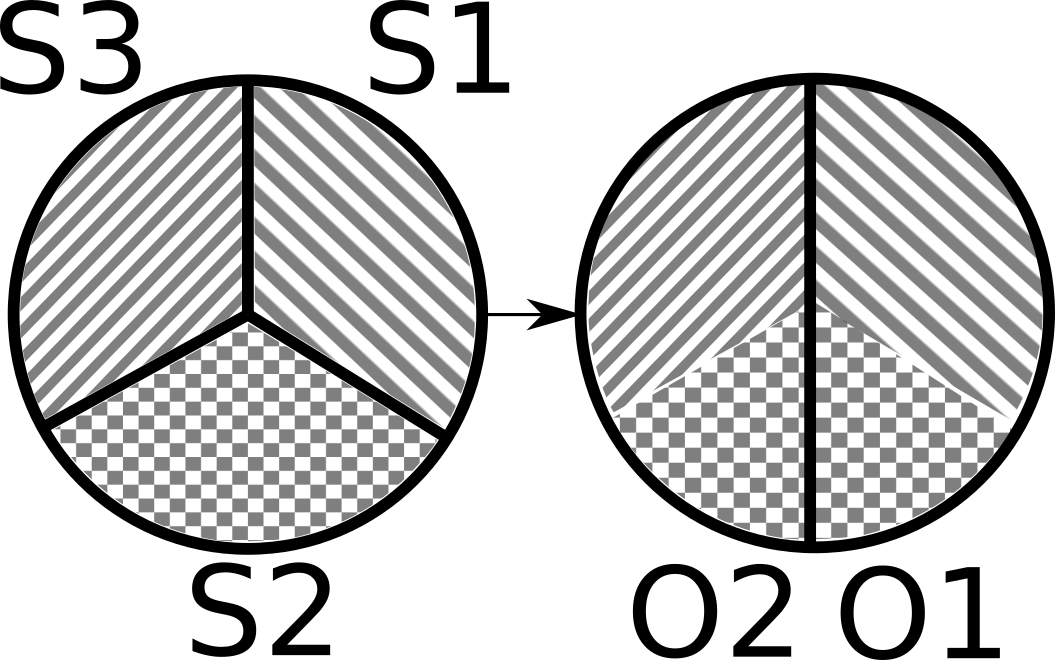}
	\caption{A translation operation between two plural nodes shared by different numbers of processes.}
	\label{fig:Translation}
\end{figure}

To understand the scope of the problem, consider Figure \ref{fig:Translation}. Here, the source node is a plural node shared by three processes (S1, S2 and S3); the observer node is shared by two processes (O1 and O2). Multipole samples for both are shared starting at the top of each circle and increasing clockwise (consistent with the process alignment scheme utilized during M2M). Process S1 and O1 both own multipole samples with the lowest indices, with S1 having less samples than O1. For this far-field interaction, S1 would need to send all samples that it owns to O1. S3 and O2 both own samples with the highest indices; here S3 would need to send all of its samples to O2. Finally, S2 owns samples that are needed by both O1 and O2, therefore S2 must send half of its samples to O1 and the other half to O2. Since each node in the H-FMM tree interacts with several others ($\approx$27 for surfaces and up to 189 for volumes) each of which may be shared by a varying number of different processes, it is evident that coordination of all communications that must take place during an H-FMM evaluation is non-trivial. 

In an initialization step before the actual M2L operations, all processes discover the owner process(es) of the tree nodes (i.e., observers) which will need the multipole data for the source nodes they own, given the partitioning of leaf nodes (for load balancing purposes) and process alignments for plural nodes. This pre-calculated list is formed to carry out the actual communications that will take place during the ensuing H-FMM potential evaluations. If a source node or the corresponding observer node is plural, then the pre-calculated communication list will include only the intersection of the multipole data owned by both the source and target processes. When a source node on a process is needed by multiple observers on another process, it is sufficient to include the source node in the communication list to that other process only once. Also, multipole data for multiple source nodes residing on one process that are needed by another can be combined into a single message in this communication list, even if the source nodes are at different levels.  
Note that the tree structure in most H-FMM applications are fixed. As such, the overheads associated with such an initialization stage is minimal. 

\begin{algorithm}
	\caption{M2L Translation}\label{M2LTrans}
	\begin{algorithmic}[1]
		\State Determine the union of data owned by both source and target.
		\Procedure{M2L}{}
		\For{each source box}
		\For{each farfield interaction target}
		\If{Target box is not local}
		\For{each process sharing target box}
		\State Add source multipole data in union of data owned by both source and target to buffer.
		\EndFor
		\EndIf
		\EndFor
		\EndFor
		\State Communicate buffers between processes.
		\For{each target box}
		\For{each farfield interaction source}
		\If{Source box is local}
		\State Load source multipole data from local memory.
		\State Translate source multipole data to target and add to existing translated data for target node
		\Else
		\For{each process sharing source box}
		\State Read source multipole data from communication.
		\State Translate source multipole data to target and add to existing translated data for target node
		\EndFor
		\EndIf
		\EndFor
		\EndFor
		\EndProcedure
	\end{algorithmic}
\end{algorithm}

While plural node to plural node far field interactions (which constitute the most expensive communications in an H-FMM evaluation) could actually be carried out using all-to-all communications that involve only the users of the two corresponding plural nodes, due to the excessive number of M2L interactions present in large-scale computations (and hence the large number of different communicators that must be created), we choose to perform these communications using non-blocking point-to-point send/recv operations (i.e., MPI-Isends and MPI-Irecvs) in the default global communicator. Another reason for opting for a point-to-point scheme is that there are significant differences in the amount of data that must be sent to one process compared to another. As part of the initialization step then, each process allocates a message buffer for every other process that it will communicate with. The size of the send buffer is limited to avoid excessive memory use and maximize communication overlap. 

To perform communications during M2L, every process first fills their send buffers for each process based on the pre-calculated communication list and initiates the message transmission using \verb|MPI_Isend|s. Immediately after initiating the sends, each process starts waiting for their expected messages using \verb|MPI_Irecv|s. The status of these communications are checked periodically. Computations are overlapped with M2L communications in two ways. First, blocks of translations that are entirely local (which is actually common at the lower tree levels) are processed while the non-blocking send/recvs are taking place. Second, translation data that is detected as received during the periodic checks are applied immediately, overlapping the corresponding computational task with communications underway. Due to the limit we impose on the message buffer sizes, communications with processes that involve a large amount of data transfer need to be performed in multiple phases. Hence, upon reception/delivery of a message from another process, if there is more data to be transferred, a new non-blocking recv/send operation is initiated. 

%During the actual translation process, the initialization allows a process to determine the start and end sample either to send for each source node the process owns, or to translate and apply for each observer node the process owns.
%, and during the initialization phase S1 and O1 will communicate that S1's communication will begin with the first sample for the node and end with the last sample S1 owns
%and during initialization S3 and O2 communicate that S1's communication begins with the first sample it owns and ends with the last sample of the node. 
%During initialization S2 will communicate with both O1 and O2. S2 will communicate with O1 that it will send samples starting with the first sample S2 owns, and the last sample O1 owns. S2 will communicate with O2 that it will send samples starting with the first sample O2 owns, and ending with the last sample S2 owns. 

\paragraph{Translation Operators}
Source node data is translated to the target node by multiplying it with a translation operator. The translation operators can be pre-calculated to reduce computational costs. As these can potentially take significant memory, we limit such memory use by each process by having them store the pre-calculated operators only for translation of the local and remote samples that they will actually need. This information can be determined from the pre-calculated M2L communication list. In case the memory available is not sufficient to store the needed operators, we use techniques outlined in \cite{hughey2019parallel} to sample and interpolate translation operators.

\subsubsection{L2L}
\label{sec:l2l}
To anterpolate and distribute the translated local expansions down to the child nodes, L2L applies the operations in M2M in reverse order. First, local expansions at the parent node are shifted to the center of each child node, they are then anterpolated and percolated down the tree. Finally, the anterpolated data is aggregated with local expansions previously translated to the child node during the M2L stage.

Much like M2M, L2L operation is parallelized in three different ways: i) Non-plural parent tree nodes at the lower tree levels are processed independently in parallel by their owner processes, ii) for a plural node with a non-plural child, shifts involve communications, but the ensuing anterpolation and aggregation (with translated local expansions) are performed sequentially by the process owning the child node, iii) plural nodes with plural children require fine-grained parallelization. 

\paragraph{Shift}
In a parallel shift operation, parent node samples corresponding to those of the child node must be communicated by the processes sharing the parent node to the process(es) owning the child node. This is most easily done before the data has been shifted, as the parent will not need to know the position of the child node. In case of plural parent and non-plural child, this communication would essentially be a \emph{gather}, and in case of plural parent with a plural child, it would be an \emph{all-to-all}, in both cases involving all processes sharing the parent node. However, only a subset of the processes sharing the parent node will actually share the child node. To avoid non-trivial issues that would arise from having to coordinate several collective calls among different subsets of processes, we again resort to point-to-point communications instead. Consequently, messages are only sent from processes owning a piece of the parent node to a process owning the corresponding piece of the child node. Once a process gathers all of necessary samples of the parent node, it applies the shift operation to all its multipole samples independently.

\paragraph{Anterpolation}
Anterpolation would have to be performed in parallel only if the child node is a plural node. The procedure for parallel anterpolation is exactly the same as that of the parallel interpolation, except that the number of multipole samples is reduced (rather than increased).

\paragraph{Aggregation}
Aggregating the shifted and anterpolated parent data with translated local expansions is trivial. Even in the case of plural child nodes, all required data is already available locally.

\subsubsection{L2O}
\label{sec:m2m}
As in C2M, each process handles the L2O computations of its assigned leaf nodes in parallel independently.

\section{Computational and Communication Complexity Analysis}

\label{sec:complexity}

In this section, we analyze the asymptotic computational and communication complexity of the parallel H-FMM algorithm described above. To simplify the analysis, we focus on two extreme cases, a 2D surface represented by points on a regularly spaced planar grid (dimension $d$=2) and a 3D volumetric structure represented by points on a regularly spaced cubic grid ($d$ = 3). These represent extreme cases, and hence are ideal for asymptotic analysis. 

Let $K(l)$ denote the number of samples in the $\theta$ and $\phi$ directions for a node at level $l$. Assume that each leaf node contains $\mathcal{O}(1)$ samples. It follows that the number of leaf nodes is $\propto N_s$, the number of source points. For simplicity and with no loss in generality, we assume that the constant of proportionality is 1. Next, we denote the number of nodes at level $l$ by $G(l)$. The total number of levels is given by $N_L$. %We choose to enumerate the leaf level as level 1 and the root level as level $N_L$ because this helps with finding simpler expressions for summations. 
As one moves up the octree, we observe that the number of groups per level is reduced by roughly 4 times for the 2D surface and 8 times for the 3D volume. Leveraging the relation between the dimensionality of a structure and the rate of decrease in the number of nodes per level,  one can write $G(l)$ as follows: 
\begin{equation}
G(l) = \frac{N_s}{(2^d)^{(l-1)}} \label{eq:gl}.
\end{equation}
% While our analyses below specifically focus on a planar surface and an ideal cube, it can potentially be extended to structures ``in between", as the average decline rate in $G(l)$ could be captured with 1/$2^d$, where $d$ would be slightly larger than but close to 2 for surface-like geometries and it would be slightly less than but still close to 3 for volume-like geometries. 
Since $K(l)$ doubles at each level, given $K(1) = C_k$, it follows that 
\begin{equation}
    K(l) = 2^{(l-1)} C_k \label{eq:kl}.
\end{equation}
Finally, we define $P$ as the number of processes, $P_L$ as the level where (almost) all nodes in a level start becoming plural and $P_N(l)$ as the average number of processes sharing a plural node at level $l$. Equivalently, $P_L$ is  the level when $G(P_L) <P $ for the first time, and this remains true from hereon to the root. Given $P$, $N_s$ and $d$,

\begin{equation}
P_N(l) = \frac{P}{G(l)} = \frac{P(2^d)^{(l-1)}}{N_s} \label{eq:pn}
\end{equation}

\subsection{Interpolation (M2M)}

\paragraph{Computational Complexity} 
M2M is performed for each node, starting from the leaf level up to the highest level $N_L$. The dominant component in the computational complexity for $M2M$ is FFT-based interpolation. Shifting and aggregation are $O(K(l)^2)$ operations each, while interpolation for a given node costs $O(K(l)^2 \log^2{(K(l)))}$. This gives a total computational complexity of 

\begin{equation}
C \propto \sum_{l=1}^{N_L}G(l)K(l)^2 \log^2 (K(l))
\end{equation}

Plugging in the equations \eqref{eq:gl} and \eqref{eq:kl} and simplifying the summation, we obtain the computational complexity for a surface to be:

\begin{equation}
C \propto O(N_s \log^2 N_s),
\end{equation}

and for a volume to be:

\begin{equation}
C \propto O(N_s)
\end{equation}

\paragraph{Number of Messages (Latency)} 
Communication in $M2M$ happens during aggregations for both coarse-grained and fine-grained parallel $M2M$s, as well as the FFTs of the fine-grained parallel $M2M$s.  As described in Sect.\,\ref{sec:m2m}, we perform aggregations (which are effectively reduce-scatter operations) using point-to-point communications. In an ideal tree, every source and observer node will be divided among the same number of nodes. This means the portion of a source node owned by any process will only be owned by a single process in the observer node, limiting the communication for each source node process to one process in each observer node. Since this is done for each group at each level, the total message count for aggregations can be written as: 
\begin{equation}
M_{Ag} \propto \sum_{P_L}^{N_L} G(l) P_N(l+1).
\end{equation}
Using expressions for $P_N(l+1)$ and $G(l)$, yields the number of messages for aggregation
\begin{equation}
M_{Ag} = O(P \log (N_S) 2^d).
\end{equation}
Here, we ignore aggregations that would be needed for plural nodes (located at process boundaries) below level $P_L$. Note that there may only be two such plural nodes per level for each process and these aggregations will involve only two processes. As such, their contribution to the number of messages during aggregations is of a lower order term. 

Next, consider the parallel FFTs in fine-grained parallel $M2M$s. In this case, an all-to-all communication is performed after each of the two fold and transpose operations. As we implement these all-to-all communications using point-to-point calls, the message count for FFTs is then:

\begin{equation}
M_{FFT} = \sum_{l=P_L}^{N_L} G(l) P_N(l)^2 \propto O(P^2).
\end{equation}
Consequently, the total number of messages for $M2M$ is:
\begin{equation}
M_{M2M} = O(P^2 + P \log(N_s)). \label{eq:m2m}
\end{equation}
Note, the $N_s$ portion of the equation above is only going to matter when $P_L$ is greater than the number of levels in the tree. In all other cases, increasing the height of the tree does not increase the number of levels with plural nodes. Given that it is practically useless to have more processes than the number of leaf nodes (which is the condition required for $P_L$ to be more than the tree height), the message count can be simplified to $M_{M2M} = O(P^2)$.

\paragraph{Communication Volume (Bandwidth)}
Bandwidth during interpolation is due to all to all communications during interpolation, and a reduce scatter during the aggregation. Each of these operation communicates up to the entire node, resulting in a bandwidth that can be written as:
\begin{equation}
B \propto \sum_{l=1}^{N_L} G(l) K(l)^2
\end{equation}
Applying the previous definitions for $G(l)$ and $K(l)$ yields a communication bandwidth of $B \propto N_s \log N_s$ for the surface geometry, and $B \propto N_s$ for the volume geometry.

\subsection{Translation (M2L)}

\paragraph{Computational Complexity}
%Performing the far-field interaction between two nodes at the same level involves translating the source node's multipole expansions to the center of the target node, and adding its impact into the target node's local expansion. This operation is facilitated by a pre-calculated translation operator which is vector that simply multiplies each multipole sample element by element. 

The complexity for the translation operation at a given level is directly proportional to the number of multipole samples for nodes, the average number of interactions per node (denoted by $I(l)$ for level $l$), and the number of nodes at that level. Summing these costs across all levels, we obtain:
\begin{equation}
C \propto \sum_{l=1}^{N_L} K(l)^2 I(l) G(l).
\label{eq:mMapping}
\end{equation}
While the number of interactions for a node changes based on its exact position in the geometry (for instance, corner or edge nodes), the upper limit is the constant $6^d-3^d$. Using the equations for $K(l)$ and $G(l)$, computational complexity of the translation step can be simplified to $O(N_s \log N_s)$ for the surface structure, and to $O(N_s)$ for the volume structure.

\paragraph{Number of Messages (Latency)}
At level $P_L$ or above, a process can have multipole samples for only one node. Since a process owns at most part of a single node, each of its interactions will require a separate communication because the nodes in its far-field will all reside on different processes. Assuming an ideal tree partitioning where the source and target nodes are shared among the same number of processes, the $k$th process for the target node will only need the source node data from the $k$th process of the source node. As we limit the size of each translation message, the number of messages will then be proportional to the communication volume between a pair of processes divided by the message buffer size $M_S$. At levels below $P_L$, a process can own multiple nodes. Here groups of nodes can be communicated to the same process, if all source nodes reside on one process and all observer nodes reside on another.  In this case, the interaction count is going to be based on the total amount of data communicated between the two interacting processes, divided by the message buffer size, summed up for all interacting processes.

Considering contributions at/above $P_L$ and below $P_L$ gives a total message count of:
\begin{equation}
M \propto \sum_{l=P_L}^{N_L} P I(l) \frac{K(l)^2}{P_N M_S} + P I(l) \ceil{\frac{\sum_{l=1}^{P_L - 1} K(l)^2 \frac{G(l)}{P}}{M_S}}
\label{eq:mMapping}
\end{equation}
where $M_S$ is the size of message buffers. For the surface geometry, this can be simplified to $M \propto O(N_s \log N_s) + O(N_S)$ (where the first term is for levels $\ge P_L$ and the second term is for levels $<P_L$), and for the volume geometry it can be simplified to $M \propto O(N_s)$ (with both below and above $P_L$ having the same impact).

\paragraph{Communication Volume (Bandwidth)}
Similarly, communication volume can be analyzed in two parts as well. At and above $P_L$, all multipole data for every source node must essentially be communicated to every target node as no process contains any multipole data other than its own. Even if the number of processes increases, still the same amount of data needs to be transmitted, just among an increased number of nodes. Therefore, for level at or above $P_L$, the communication volume is independent of the number of processes:
\begin{equation}
B \propto \sum_{l=P_L}^{N_L} K(l)^2 G(l) I(l)
\end{equation}
This expression simplifies to $O(N_s \log N_s)$ for the surface geometry and to $O(N_s)$ for the volume geometry.

Below $P_L$, each process will own more than one node, nodes will be interacting with nodes on the same process, or multiple nodes owned by a neighboring process. In fact, only nodes within two nodes off the edge of process boundaries will require communications with other processes. Total communication bandwidth can then be expressed as:
\begin{equation}
B \propto \sum_{l=1}^{P_L - 1}P K(l)^2 (S_N)
\end{equation}
%\begin{equation}
%B \propto \sum_{l=SplitLev + 1(2?)}^{N_L}P K(l)^2 (12L_N + 36)
%\end{equation}
where $S_N$ is the number of nodes that have nodes in its far-field from at least one (out of the 8 possible neighboring processes for the surface and 26 for the volume) other processes touching them and can be represented as $S_N = \frac{G(l)}{P}^{\frac{d-1}{d}} = \frac{1}{P}(\frac{N_s}{(2^d)^{(l-1)}})^{\frac{d-1}{d}}$.
With this definition of $S_N$, the total communication volume for M2L below $P_L$ becomes
\begin{equation}
B \propto O(N_S)
\end{equation}
for the surface, and
\begin{equation}
B \propto O(N_S^{\frac{2}{3}})
\end{equation}
for the volume, due to the lower portion of the tree dominating. 

\subsection{Anterpolation (L2L)}

As mentioned before, L2L is the reverse operation for M2M. Similar to M2M, anterpolation dominates the computational complexity for L2L. Computational complexity for anterpolations is the same as that of interpolations, so L2L's computation complexity is the same as M2M's. 
%In M2M, interpolated data is shifted, aggregated and then distributed to the processes sharing the parent node. In L2L, the parent node data must be communicated from each process owning the portion of the parent node that the process owning the corresponding portion of the child node will use to shift from the parent node center to the child node center and then anterpolate. 
Likewise, communications performed are the same but in reverse order. Therefore, the latency and bandwidth costs of L2L are the same as those of M2M.

%\paragraph{M2M/L2L}
%\paragraph{Computational Complexity}

\section{Performance Evaluation}

In this section, we evaluate the performance of the parallel H-FMM algorithm described. All results were obtained on the Cori-Haswell supercomputer at National Energy Research Scientific Computing Center (NERSC). Each node on this system contains two sockets, populated by Intel Xeon E5-2698 v3 (Haswell)  processors  with  a  clock  speed  of  2.3\,GHz. Each  node  has  32  cores  and  128 GB 2133MHz DDR4 RAM. The code is implemented in Fortran 90 using only MPI parallelization and was compiled with the Intel compiler. The FFTW library is used for all FFT operations.

The runs here focus on the timing of the M2M, M2L and L2L phases of the tree traversal.  As such, each leaf node is only populated with a single unknown, effectively bypassing the near-field, C2M and L2O processing steps. This also makes the number of unknowns being processed much smaller than what could be processed by M2M, M2L and L2L in the same amount of time. In a typical $0.25\lambda$ leaf box (as we use in the runs below) with a $0.1\lambda$ discretezation rate, the number of Rao-Wilton-Glisson (RWG) functions will be around 20-30. The number of particles per box to simulate 20-30 RWG functions can range from 100-180 particle to box. Our largest tree being processed is 14 levels with 42 million points for one point per leaf box. If the leaf nodes were fully populated, this tree would be equivalent to processing a tree with 4.2 to 7.5 billion points. Populating leaf nodes would increase the time of the C2M and L2O steps, but have no impact on the execution times of the M2M, M2L and L2L phases.

\subsection{Load balance with the fine grain parallel algorithm}

The intent of the fine-grain parallel algorithm is to provide improved balance at the upper levels of the tree where a lower number of much larger nodes reside. First, we look at the performance of a planar grid of particles (in the $z=0$ plane) of dimensions $512\lambda\times512\lambda$  with a grid spacing of $\lambda/4$ and 4,194,304 particles in total. The box size is chosen to be $0.25\lambda$, resulting in a 12-level tree with 10 levels of computation. As can be seen in the left subfigure of Fig.\,\ref{fig:computational_complexity}, the resulting execution profile is very balanced across process ranks. Execution time of the fastest to the slowest process varies by only 1.43\%. Balance of total time can be a little misleading as M2L cannot progress until all processes that a given process interacts with have completed their M2M processing. However, the M2M execution times are also very balanced, varying from slowest to fastest process by 5.23\%.

\begin{figure}[H]
\centering
\includegraphics[width=.5\textwidth]{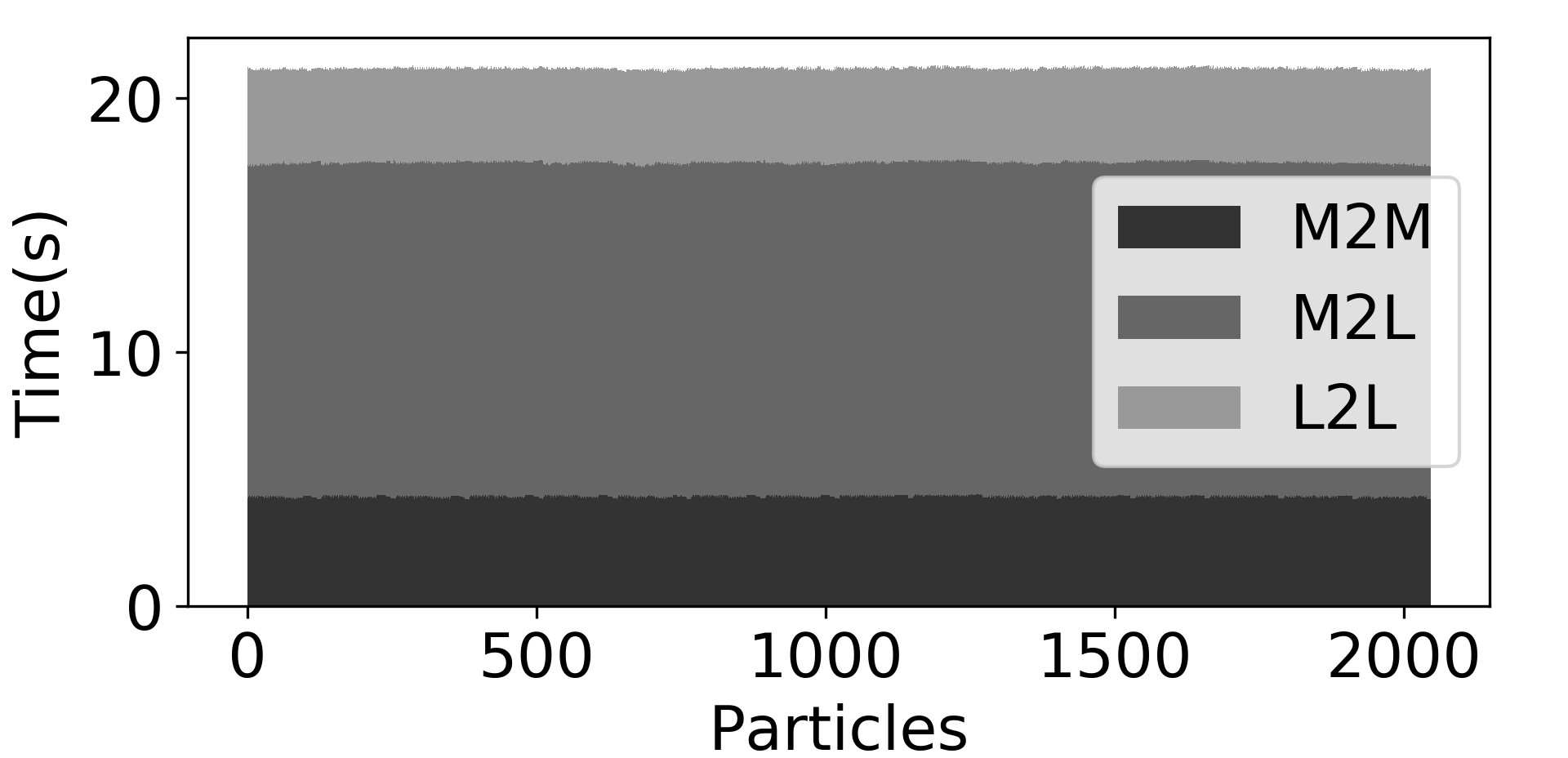}\hfill
\includegraphics[width=.5\textwidth]{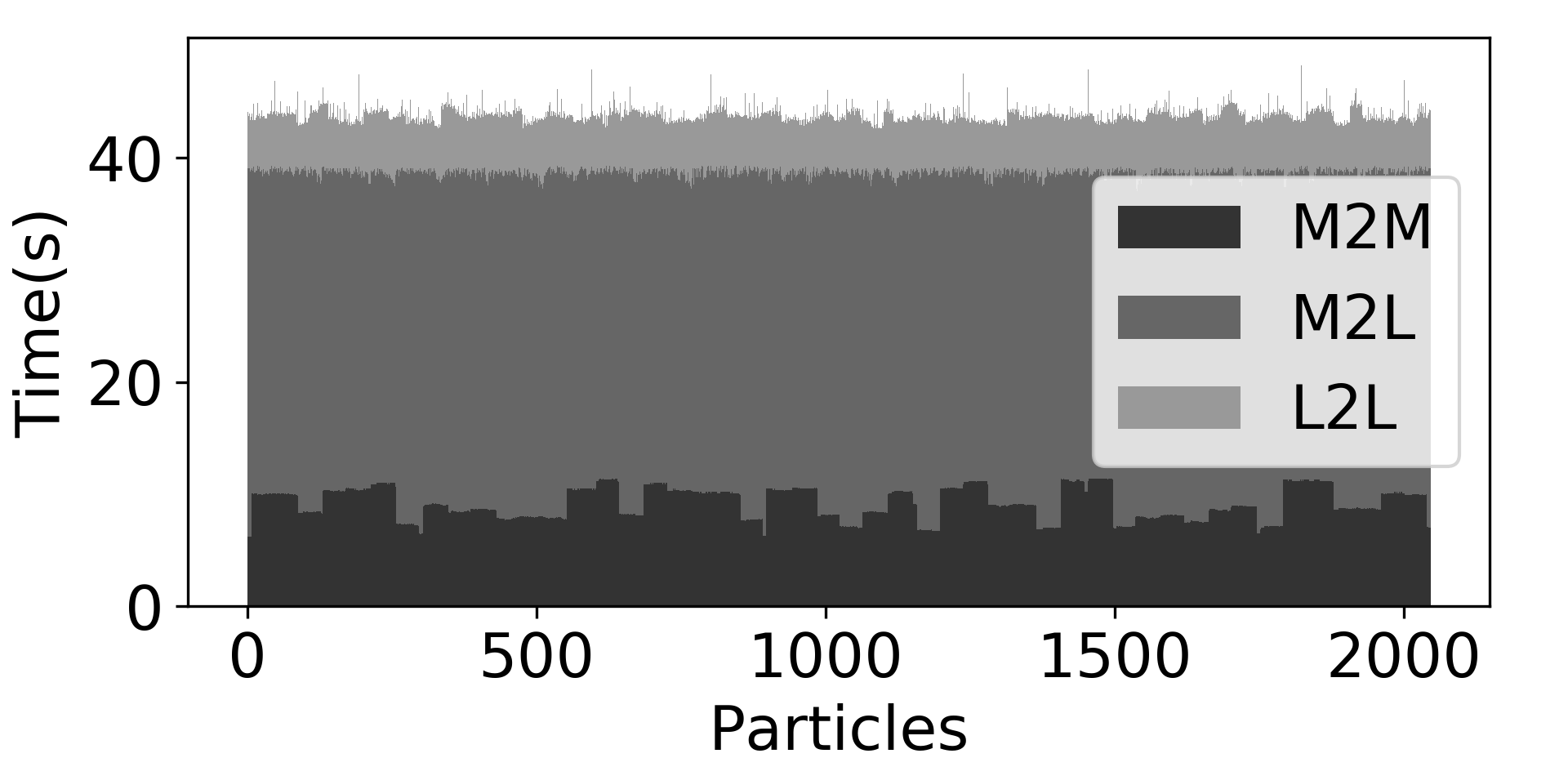}\hfill
\caption{Process execution times for a grid and a sphere geometry.}
\label{fig:computational_complexity}
\end{figure}

Next, we look at the performance for a sphere of diameter $384\lambda$ discretized using 4,542,208 dipoles on the surface with a leaf box size of $d_0=0.25\lambda$, yielding an 11-level tree. This geometry is less balanced than the grid geometry (see the right subfigure of Fig.\,\ref{fig:computational_complexity}) as high level nodes can range from having no children due to no particles being in that part of the geometry at the leaf level, up to having a completely filled quad tree from the leaf level up to a high level node. This results in notable imbalance at the M2M level, which as discussed before, results in delays in the M2L execution. Note that there is an implicit barrier at the beginning of M2L processing, as an M2L interaction communication cannot proceed until both interacting processes have completed their M2M phases (though the faster process can perform local translations while waiting). Another (less significant) implicit barrier occurs at the beginning of L2L where nodes that are fully owned by a process must have all of the data from the translated parent node to perform anterpolation on this parent node. The M2M execution range from the fastest to the slowest processes varies by as much as 84.5\%. Despite this noticeable imbalance, the long execution times are not clustered among a small group of processes as one would see if each of the highest level tree nodes were to be handled by a single process without fine-grain parallelization. So even in this unbalanced geometry, the fine-grain parallel algorithm is helping to maintain a good load balance across process ranks.

\subsection{Scalability}

Next, we investigate the strong scaling efficiency of our parallel Helmholtz FMM algorithm, first on a surface and then on a volumetric structure.
%An analysis of the parallel efficiency (\emph{strong scaling}), defined as
%\begin{equation}
%\text{Eff.}\left[N_q\right] (\%) \doteq \frac{N_p T_{N_p}}{N_q T_{N_q}} \times 100 \%,
%\end{equation}
%where $N_p,T_{N_p}$ are the reference process count and computation time, and $N_q,T_{N_q}$ are the process count and computation time for which the efficiency is being calculated, shows where the parallel execution is performing well and where it is performing more poorly. 

For the 2D surface structure, we use the same $512\lambda\times512\lambda$ planar grid as above. As our base case for strong scaling efficiency, we use the performance on 128 cores because this is the smallest number of cores that this problem can be executed on due to its memory requirements. As seen in Table\,\ref{fig:pfft_grid_timings_table}, both the interpolation and anterpolation phases (M2M and L2L) perform very well with the increasing process counts, while M2L's performance falls off rapidly (down to 25\% efficiency on 2048 cores). There are a couple of factors that contribute to this difference we observe in scaling characteristics. First factor is that M2M and L2L incur significant communications only at the highest level nodes, while M2L communications occur at every level where the source and observer nodes are on separate processes, which may essentially happen all the way down to the leaf nodes. Secondly, and more importantly, M2M and L2L computations involve relatively computation-heavy FFTs in between its communication steps. When the number of nodes in a level exceeds the number of processes, no process can own both a source and observer node of any translation, so all node data must be communicated. As the number of processes approaches the number of leaf nodes, the M2L communication bandwidth asymptotically approaches the worst case estimate. This means the increase in M2L bandwidth exceeds the worst case estimate increase as the number of processes approaches the number of leaf nodes.
%the worst case M2L bandwidth estimate shows a only dependent on the number of samples, but this is not the full story. Once there are enough processes that every leaf is on a separate process, increasing the number of processes would not increase M2L bandwidth. For a 12 level tree of a surface geometry, this would require 4,194,304 processes. So as the number of processes increases toward the number of leaf nodes, the M2L bandwidth trends toward the worst case bandwidth. Support for this can be seen by noting that while parallel efficiency drops off drastically from 128 to 256 processes, but less so from 1024 to 2048 processes. 
Despite M2L not scaling very well, the fine grained parallel algorithm presented still provides good speedups, nearly an 8x speedup when going from 128 to 2048 processes without showing any performance stagnation.

\begin{table*}
\centering
%\rowcolors{2}{gray!30}{white}
\begin{tabular}{|r|r|r|r|r|r|r|r|r|r|r|}
\multicolumn{1}{c}{} & \multicolumn{4}{c}{\textbf{Grid} (s)} & \multicolumn{1}{c}{\textbf{Speedup}} & \multicolumn{4}{c}{\textbf{Par Eff.} (\%)}\\\hline
$N_p$ & M2M & M2L & L2L & Tot & Tot & M2M & M2L & L2L & Tot\\\hline
128 & 5.80 & 5.30 & 5.30 & 18.55 & 1.00 & 1.00 & 1.00 & 1.00 & 1.00\\
256 & 3.06 & 4.13 & 2.66 & 11.21 & 1.65 & 0.95 & 0.64 & 0.99 & 0.83\\
512 & 1.52 & 2.76 & 1.31 & 6.42 & 2.89 & 0.95 & 0.48 & 1.01 & 0.72\\
1024 & 0.81 & 2.12 & 0.69 & 4.05 & 4.58 & 0.89 & 0.31 & 0.95 & 0.57\\
2048 & 0.43 & 1.31 & 0.37 & 2.38 & 7.78 & 0.84 & 0.25 & 0.89 & 0.49\\\hline
\end{tabular}
\caption{Performance of the MLFMA algorithm on the $512\lambda$ grid geometry.}
\label{fig:pfft_grid_timings_table}
\end{table*}

Next, we examine strong scaling on a $32\lambda\times32\lambda\times32\lambda$ volumetric structure (Table\,\ref{fig:pfft_vol_timings_table}). From 128 to 512 processes, we observe very good scaling (80\% overall efficiency), but then parallel efficiency drops off quickly (down to 50\% overall at 2048 cores). In a volumetric problem, each tree node has a large number of nodes in its far-field (up to 189). Therefore the overall execution time is largely dominated by the M2L stage which does not manifest good scaling. The ideal scenario for our fine-grained parallel algorithm is when the nodes of a given level are distributed evenly among the processes, i.e., when the number of processes divides evenly into the number of nodes in a level or vice versa. This does not occur at 1024 or 2048 processes for this particular volumetric problem. Nevertheless, the overall speedup remains at around 8x when going from 128 to 2048 processes.
%, so the parallel execution is still providing value.

\begin{table*}
\centering
%\rowcolors{2}{gray!30}{white}
\begin{tabular}{|r|r|r|r|r|r|r|r|r|r|r|}
\multicolumn{1}{c}{} & \multicolumn{4}{c}{\textbf{Volume} (s)} & \multicolumn{1}{c}{\textbf{Speedup}} & \multicolumn{4}{c}{\textbf{Par Eff.} (\%)}\\\hline
$N_p$ & M2M & M2L & L2L & Tot & Tot & M2M & M2L & L2L & Tot\\\hline
128 & 0.527 & 2.15 & 0.526 & 3.26 & 1.00 & 1.00 & 1.00 & 1.00 & 1.00\\
256 & 0.266 & 1.10 & 0.263 & 1.68 & 1.94 & 0.99 & 0.97 & 0.99 & 0.97\\
512 & 0.14 & 0.679 & 0.135 & 0.99 & 3.27 & 0.93 & 0.79 & 0.97 & 0.82\\
1024 & 0.079 & 0.380 & 0.084 & 0.574 & 5.68 & 0.83 & 0.71 & 0.78 & 0.71\\
2048 & 0.051 & 0.271 & 0.058 & 0.406 & 8.03 & 0.65 & 0.50 & 0.57 & 0.50\\\hline
\end{tabular}
\caption{Performance of the MLFMA algorithm on the $32\lambda$ volumetric geometry.}
\label{fig:pfft_vol_timings_table}
\end{table*}

Finally, we look at scaling on the $384\lambda$ sphere (Table\,\ref{fig:pfft_sphere_timings_table}). As seen in the load balance analysis of the previous subsection, load imbalances result in the faster processes having to wait for slower processes. This results in a noticeable drop in scaling efficiency of the M2M phase, where the imbalance has the greatest impact, as well as the M2L phase, where some processes that are already in their M2L phase have to wait for others that are still in their M2M phase. This also has an impact on the overall speedup. While increasing the number of processes continues to improve execution times, the speedup when going from 128 to 2048 processes is just under 5x in the sphere case.

\begin{table*}
\centering
%\rowcolors{2}{gray!30}{white}
\begin{tabular}{|r|r|r|r|r|r|r|r|r|r|r|}
\multicolumn{1}{c}{} & \multicolumn{4}{c}{\textbf{Sphere} (s)} & \multicolumn{1}{c}{\textbf{Speedup}} & \multicolumn{4}{c}{\textbf{Par Eff.} (\%)}\\\hline
$N_p$ & M2M & M2L & L2L & Tot & Tot & M2M & M2L & L2L & Tot\\\hline
128 & 6.71 & 13.54 & 5.29 & 26.76 & 1.00 & 1.00 & 1.00 & 1.00 & 1.00\\
256 & 3.86 & 10.05 & 2.7 & 18.00 & 1.49 & 0.87 & 0.67 & 0.97 & 0.74\\
512 & 2.34 & 6.20 & 1.46 & 11.04 & 2.42 & 0.72 & 0.55 & 0.91 & 0.61\\
1024 & 1.23 & 4.32 & 0.72 & 7.70 & 3.47 & 0.68 & 0.39 & 0.92 & 0.43\\
2048 & 0.92 & 3.19 & 0.416 & 5.58 & 4.79 & 0.46 & 0.26 & 0.79 & 0.30\\\hline
\end{tabular}
\caption{Performance of the MLFMA algorithm on the $384\lambda$ diameter sphere geometry.}
\label{fig:pfft_sphere_timings_table}
\end{table*}

\subsection{Complexity Analysis}

To help validate the complexity analysis presented in Sect.\,\ref{sec:complexity}, the software was instrumented to report the computational cost, the number of  messages sent and the size of these messages. In accordance with the geometries analyzed in Sect.\,\ref{sec:complexity}, data was collected on the grid geometries ranging from $64\lambda$ to $1024\lambda$ and volume geometries ranging from $16\lambda$ to $16\lambda$ to as these geometries produce perfect quadtrees of heights ranging from 9 to 13 levels and octrees of heights ranging from 7 to 11 levels, respectively. As complexity estimates are asymptotic, they are scaled by least-squares fit to help visualize how well the estimates match the actual measurements.

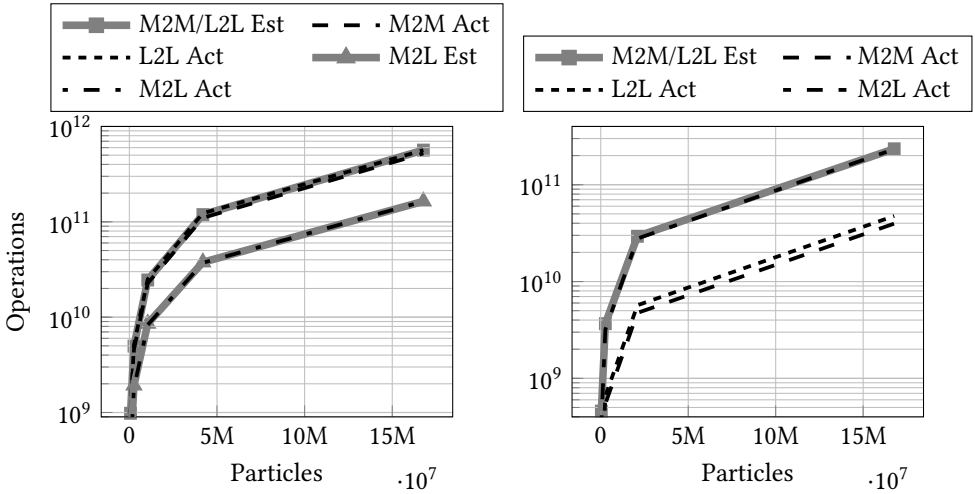
\begin{figure}
    \centering
  \begin{tikzpicture}
\begin{semilogyaxis}[cycle list/Set1, % list of colors for plotting data
            grid=both, % sets the background grid
            xlabel={Particles},
            ylabel={Operations},
            ytick pos=left,
            xtick pos=left,
            xtick={0,5000000,10000000,15000000},
            xticklabels={0,5M,10M,15M},
            ymin=900000000,ymax=1000000000000,
            legend style = {legend cell align=left,/tikz/every even column/.append style={column sep=0.25cm},at={(0.01,0.01)},anchor=south west},
        legend image post style={scale=1.5},
        legend style={at={(0.5,1.05)},anchor=south}, % place the legend slightly above the plot
        legend columns=2, % two columns in the legend
        set layers,mark layer=axis background,
        width=0.45\textwidth
            ]
  \addplot+[gray, line width=1mm, style={mark size=1.0},mark size=1.0,mark=square*,mark options=solid]
           table[scatter, x index={0},y index={1}]
           {\gridcomplex};
  \addplot+[black,dash pattern=on 6pt off 6pt on 6pt off 6pt, line width=0.5mm]
           table[scatter, x index={0},y index={2}]
           {\gridcomplex};
  \addplot+[black,dash pattern=on 3pt off 3pt on 3pt off 3pt, line width=0.5mm]
           table[scatter, x index={0},y index={3}]
           {\gridcomplex};
  \addplot+[gray, line width=1mm, style={mark size=1.0},mark size=1.0,mark=triangle*,mark options=solid]
           table[scatter, x index={0},y index={4}]
           {\gridcomplex};
  \addplot+[black,dash pattern=on 2pt off 6pt on 6pt off 6pt, line width=0.5mm]
           table[scatter, x index={0},y index={5}]
           {\gridcomplex};
  % add legend captions in the order in which the \addplot's were called
  \legend{M2M/L2L Est,M2M Act,L2L Act,M2L Est,M2L Act};
\end{semilogyaxis}
\end{tikzpicture}
\begin{tikzpicture}
\begin{semilogyaxis}[cycle list/Set1, % list of colors for plotting data
            grid=both, % sets the background grid
            xlabel={Particles},
            ytick pos=left,
            xtick pos=left,
            xtick={0,5000000,10000000,15000000},
            xticklabels={0,5M,10M,15M},
            ymin=400000000,ymax=400000000000,
            legend style = {legend cell align=left,
        /tikz/every even column/.append style={column sep=0.25cm},at={(0.01,0.01)},anchor=south west},
        legend image post style={scale=1.5},
        legend style={at={(0.5,1.05)},anchor=south}, % place the legend slightly above the plot
        legend columns=2, % two columns in the legend
        set layers,mark layer=axis background,
        width=0.45\textwidth
            ]
  \addplot+[gray, line width=1mm, style={mark size=1.0},mark size=1.0,mark=square*,mark options=solid]
           table[scatter, x index={0},y index={1}]
           {\volumecomplex};
  \addplot+[black,dash pattern=on 6pt off 6pt on 6pt off 6pt, line width=0.5mm]
           table[scatter, x index={0},y index={2}]
           {\volumecomplex};
  \addplot+[black,dash pattern=on 3pt off 3pt on 3pt off 3pt, line width=0.5mm]
           table[scatter, x index={0},y index={3}]
           {\volumecomplex};
  \addplot+[black,dash pattern=on 3pt off 6pt on 6pt off 6pt, line width=0.5mm]
           table[scatter, x index={0},y index={4}]
           {\volumecomplex};
  % add legend captions in the order in which the \addplot's were called
  \legend{M2M/L2L Est,M2M Act,L2L Act,M2L Act};
\end{semilogyaxis}
\end{tikzpicture}
\caption{Actual vs. estimated computational complexity. The left subfigure shows results for the surface geometry, while the right subfigure is for the volume geometry.}
\label{fig:computational_complexity}
\end{figure}

%\begin{figure}[H]
%\centering
%\includegraphics[width=.5\textwidth]{figs/ComplexitySurface.png}\hfill
%\includegraphics[width=.5\textwidth]{figs/ComplexityVolume.png}\hfill
%\caption{Actual vs. estimated computational complexity. The left subfigure shows results for the surface geometry, while the right subfigure is for the volume geometry.}
%\label{fig:computational_complexity}
%\end{figure}

Figure \ref{fig:computational_complexity} shows the actual vs. the estimated overall computational complexities for the surface and volume geometries. The actual complexities match the estimates very closely. This indicates that the implementation of this algorithm does not have any unnecessary overhead costs in computation as computation is near to the ideal for Helmholtz FMM.

\begin{figure}
    \centering
  \begin{tikzpicture}
\begin{semilogyaxis}[cycle list/Set1, % list of colors for plotting data
            grid=both, % sets the background grid
            xlabel={Particles},
            ylabel={Bandwidth},
            ytick pos=left,
            xtick pos=left,
            xtick={0,5000000,10000000,15000000},
            xticklabels={0,5M,10M,15M},
            ymin=59000000,ymax=65000000000,
            legend style = {legend cell align=left,/tikz/every even column/.append style={column sep=0.25cm},at={(0.01,0.01)},anchor=south west},
        legend image post style={scale=1.5},
        legend style={at={(0.5,1.05)},anchor=south}, % place the legend slightly above the plot
        legend columns=2, % two columns in the legend
        set layers,mark layer=axis background,
        width=0.50\textwidth
            ]
  \addplot+[gray, line width=1mm, style={mark size=1.0},mark size=1.0,mark=square*,mark options=solid]
           table[scatter, x index={0},y index={1}]
           {\gridbandwidth};
  \addplot+[black,dash pattern=on 6pt off 6pt on 6pt off 6pt, line width=0.5mm]
           table[scatter, x index={0},y index={2}]
           {\gridbandwidth};
  \addplot+[black,dash pattern=on 3pt off 3pt on 3pt off 3pt, line width=0.5mm]
           table[scatter, x index={0},y index={3}]
           {\gridbandwidth};
  \addplot+[gray, line width=1mm, style={mark size=1.0},mark size=1.0,mark=triangle*,mark options=solid]
           table[scatter, x index={0},y index={4}]
           {\gridbandwidth};
  \addplot+[black,dash pattern=on 2pt off 6pt on 6pt off 6pt, line width=0.5mm]
           table[scatter, x index={0},y index={5}]
           {\gridbandwidth};
  % add legend captions in the order in which the \addplot's were called
  \legend{M2M/L2L Est,M2M Act,L2L Act,M2L Est,M2L Act};
\end{semilogyaxis}
\end{tikzpicture}
\begin{tikzpicture}
\begin{semilogyaxis}[cycle list/Set1, % list of colors for plotting data
            grid=both, % sets the background grid
            xlabel={Particles},
            ytick pos=left,
            xtick pos=left,
            xtick={0,5000000,10000000,15000000},
            xticklabels={0,5M,10M,15M},
            ymin=1800000,ymax=9000000000,
            legend style = {legend cell align=left,
        /tikz/every even column/.append style={column sep=0.25cm},at={(0.01,0.01)},anchor=south west},
        legend image post style={scale=1.5},
        legend style={at={(0.5,1.05)},anchor=south}, % place the legend slightly above the plot
        legend columns=2, % two columns in the legend
        set layers,mark layer=axis background,
        width=0.50\textwidth
            ]
  \addplot+[gray, line width=1mm, style={mark size=1.0},mark size=1.0,mark=square*,mark options=solid]
           table[scatter, x index={0},y index={1}]
           {\volumebandwidth};
  \addplot+[black,dash pattern=on 6pt off 6pt on 6pt off 6pt, line width=0.5mm]
           table[scatter, x index={0},y index={2}]
           {\volumebandwidth};
  \addplot+[black,dash pattern=on 3pt off 3pt on 3pt off 3pt, line width=0.5mm]
           table[scatter, x index={0},y index={3}]
           {\volumebandwidth};
  \addplot+[gray, line width=1mm, style={mark size=1.0},mark size=1.0,mark=triangle*,mark options=solid]
           table[scatter, x index={0},y index={5}]
           {\volumebandwidth};
  \addplot+[black,dash pattern=on 3pt off 6pt on 6pt off 6pt, line width=0.5mm]
           table[scatter, x index={0},y index={5}]
           {\volumebandwidth};
  % add legend captions in the order in which the \addplot's were called
  \legend{M2M/L2L Est,M2M Act,L2L Act,M2L Est,M2L Act};
\end{semilogyaxis}
\end{tikzpicture}
\caption{Actual vs estimated communication volume. The left subfigure shows results for the surface geometry, while the right subfigure is for the volume geometry.}
\label{fig:m2m_l2l_bandwidth}
\end{figure}
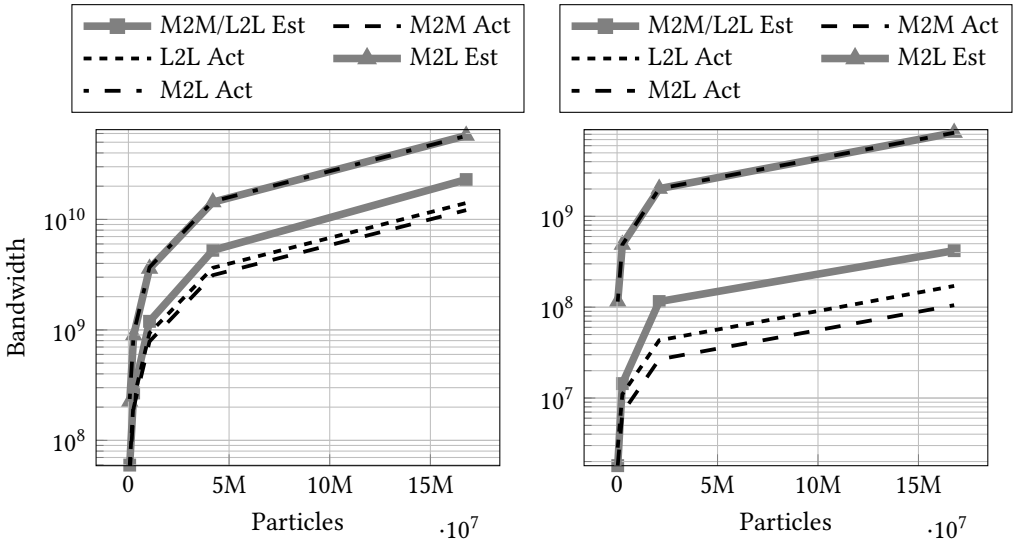

%\begin{figure}[H]
%\centering
%\includegraphics[width=.5\textwidth]{figs/M2ML2LBandwidthSurface.png}\hfill
%\includegraphics[width=.5\textwidth]{figs/M2ML2LBandwidthVolume.png}\hfill
%\caption{Actual vs estimated communication volume. The left subfigure shows results for the surface geometry, while the right subfigure is for the volume geometry.}
%\label{fig:m2m_l2l_bandwidth}
%\end{figure}

Figure \ref{fig:m2m_l2l_bandwidth} shows the actual vs the estimated communication volumes for each phase separately. Of note is how the measured communication volume drops off relative to the estimate. We believe this is due to the number of samples producing a tree with more nodes at lower levels than the number of processes. Hence, many nodes are fully owned by a single process and require no communication during M2M or L2L. Increasing the number of processes would lead to more levels with plural nodes, bringing the communication volume closer to our estimates. M2L does not show the same communication volume falloff as M2M and L2L compared to the estimated volume because fully owned nodes still require data from the source nodes to be communicated to the process owning the observer node. Such communications will be required all the way down to the leaf nodes.

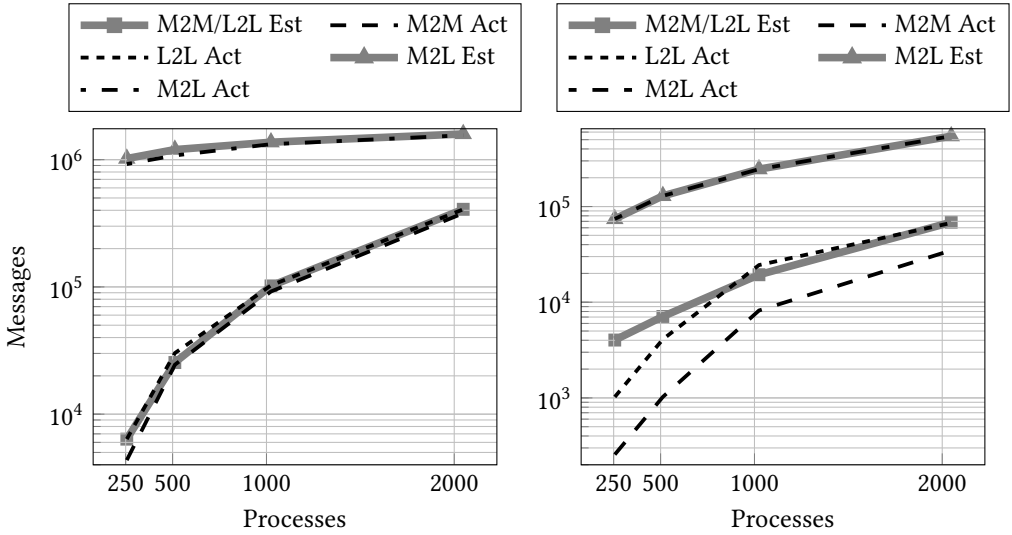
\begin{figure}
    \centering
  \begin{tikzpicture}
\begin{semilogyaxis}[cycle list/Set1, % list of colors for plotting data
            grid=both, % sets the background grid
            xlabel={Processes},
            ylabel={Messages},
            ytick pos=left,
            xtick pos=left,
            xtick={0,250,500,1000,2000},
            xticklabels={0,250,500,1000,2000},
            ymin=4000,ymax=1750000,
            legend style = {legend cell align=left,/tikz/every even column/.append style={column sep=0.25cm},at={(0.01,0.01)},anchor=south west},
        legend image post style={scale=1.5},
        legend style={at={(0.5,1.05)},anchor=south}, % place the legend slightly above the plot
        legend columns=2, % two columns in the legend
        set layers,mark layer=axis background,
        width=0.50\textwidth
            ]
  \addplot+[gray, line width=1mm, style={mark size=1.0},mark size=1.0,mark=square*,mark options=solid]
           table[scatter, x index={0},y index={1}]
           {\gridmessages};
  \addplot+[black,dash pattern=on 6pt off 6pt on 6pt off 6pt, line width=0.5mm]
           table[scatter, x index={0},y index={2}]
           {\gridmessages};
  \addplot+[black,dash pattern=on 3pt off 3pt on 3pt off 3pt, line width=0.5mm]
           table[scatter, x index={0},y index={3}]
           {\gridmessages};
  \addplot+[gray, line width=1mm, style={mark size=1.0},mark size=1.0,mark=triangle*,mark options=solid]
           table[scatter, x index={0},y index={4}]
           {\gridmessages};
  \addplot+[black,dash pattern=on 2pt off 6pt on 6pt off 6pt, line width=0.5mm]
           table[scatter, x index={0},y index={5}]
           {\gridmessages};
  % add legend captions in the order in which the \addplot's were called
  \legend{M2M/L2L Est,M2M Act,L2L Act,M2L Est,M2L Act};
\end{semilogyaxis}
\end{tikzpicture}
\begin{tikzpicture}
\begin{semilogyaxis}[cycle list/Set1, % list of colors for plotting data
            grid=both, % sets the background grid
            xlabel={Processes},
            ytick pos=left,
            xtick pos=left,
            xtick={0,250,500,1000,2000},
            xticklabels={0,250,500,1000,2000},
            ymin=200,ymax=650000,
            legend style = {legend cell align=left,
        /tikz/every even column/.append style={column sep=0.25cm},at={(0.01,0.01)},anchor=south west},
        legend image post style={scale=1.5},
        legend style={at={(0.5,1.05)},anchor=south}, % place the legend slightly above the plot
        legend columns=2, % two columns in the legend
        set layers,mark layer=axis background,
        width=0.50\textwidth
            ]
  \addplot+[gray, line width=1mm, style={mark size=1.0},mark size=1.0,mark=square*,mark options=solid]
           table[scatter, x index={0},y index={1}]
           {\volumemessages};
  \addplot+[black,dash pattern=on 6pt off 6pt on 6pt off 6pt, line width=0.5mm]
           table[scatter, x index={0},y index={2}]
           {\volumemessages};
  \addplot+[black,dash pattern=on 3pt off 3pt on 3pt off 3pt, line width=0.5mm]
           table[scatter, x index={0},y index={3}]
           {\volumemessages};
  \addplot+[gray, line width=1mm, style={mark size=1.0},mark size=1.0,mark=triangle*,mark options=solid]
           table[scatter, x index={0},y index={5}]
           {\volumemessages};
  \addplot+[black,dash pattern=on 3pt off 6pt on 6pt off 6pt, line width=0.5mm]
           table[scatter, x index={0},y index={5}]
           {\volumemessages};
  % add legend captions in the order in which the \addplot's were called
  \legend{M2M/L2L Est,M2M Act,L2L Act,M2L Est,M2L Act};
\end{semilogyaxis}
\end{tikzpicture}
\caption{Message counts vs expected Big-O message counts. The left diagram shows analysis of a surface geometry, while the right diagram shows analysis of a volume geometry.}
\label{fig:m2m_l2l_messages}
\end{figure}

%\begin{figure}[H]
%\centering
%\includegraphics[width=.5\textwidth]{figs/M2ML2LMessagesGrid.png}\hfill
%\includegraphics[width=.5\textwidth]{figs/M2ML2LMessagesVolume.png}\hfill
%\caption{Message counts vs expected Big-O message counts. The left diagram shows analysis of a surface geometry, while the right diagram shows analysis of a volume geometry.}
%\label{fig:m2m_l2l_messages}
%\vspace{-0.15in}
%\end{figure}

Figure \ref{fig:m2m_l2l_messages} shows the measured worst case messages vs the estimated worst case messages for M2M and L2L. M2M and L2L Message counts are dominated by the $P^2$ complexity of the all to all communications and the actual message count reflects this. The M2L prediction simplifies a very complex process that results in the number of M2L messages that are communicated.

%\begin{figure}[H]
%\centering
%\includegraphics[width=.5\textwidth]{figs/M2LMessageSampleComplexityGrid.png}%\hfill
%\includegraphics[width=.5\textwidth]{figs/M2LMessageSampleComplexityVolume.png}\hfill
%\caption{M2L Messages vs expected Big-O messages for changes in particle counts.}
%\label{fig:m2l_messages_sample}
%\vspace{-0.15in}
%\end{figure}
%
%Figure \ref{fig:m2l_messages_sample} shows how the actual number of M2L messages compares with the expected number. This analysis is a bit skewed as we limit the size of the communication buffer in order to save memory. Thus more messages are sent than would be necessary if all source nodes with observer nodes in a given process could be communicated in a single message, but this message is split up into multiples.

\subsection{Process Alignment}

In Table\,\ref{fig:process_alignment_table}, we compare the number of packets sent between Rank Ordered and Process Aligned schemes during M2M and L2L phases for the $512\lambda$ grid geometry. We observe a notable reduction in the number of messages exchanged, and hence the overall bandwidth, for lower process counts and continued reduction at higher process counts as expected. This reduction is likely to be effective in the relatively good scaling characteristics of M2M and L2L phases.

\begin{table*}
\centering
%\rowcolors{2}{gray!30}{white}
\begin{tabular}{|r|r|r|r|r|r|r|}
\hline
{\bf $N_P$} &  & {\bf 128} & {\bf 256} & {\bf 512} & {\bf 1024} & {\bf 2048}\\\hline
{\bf Rank Ordered} & M2M Bandwidth & 1714 & 1872 & 3009 & 3126 & 4246\\
 & L2L Bandwidth & 1206 & 1166 & 2429 & 2382 & 3649\\
 & Combined Bandwidth & 2920 & 3038 & 5438 & 5508 & 7895\\\hline
{\bf Process Aligned} & M2M Bandwidth & 1468 & 1662 & 2711 & 3272 & 4412\\
 & L2L Bandwidth & 960 & 955 & 2131 & 2104 & 3338\\
 & Combined Bandwidth & 2428 & 2617 & 4842 & 5376 & 7750\\\hline
{\bf Delta} & & -492 & -421 & -596 & -132 & -145\\\hline
\end{tabular}
\caption{Comparison of the number of packets sent between Rank Ordered and Process Aligned schemes for the $512\lambda$ grid geometry in millions of packets sent.}
\label{fig:process_alignment_table}
\end{table*}

\subsection{Memory Utilization}

\begin{table*}
\centering
%\rowcolors{2}{gray!30}{white}
\begin{tabular}{|r|r|r|r|r|r|r|r|r|r|r|}
\hline
%$N_p$ & Tdir & S/R Buffs & Trans Ops & Tree Mem\\\hline
{\bf $N_p$} & {\bf S/R Buffs} & {\bf Trans Ops} & {\bf Tree Mem}\\\hline
%128 & 34.352 & 1.652 & 107.479 & 52.741\\
128 & 1.652 & 107.479 & 52.741\\
%256 & 68.704 & 4.487 & 141.006 & 61.953\\
256 & 4.487 & 141.006 & 61.953\\
%512 & 137.409 & 102.661 & 162.204 & 52.768\\
512 & 102.661 & 162.204 & 52.768\\
%1024 & 274.818 & 256.982 & 199.047 & 62.028\\
1024 & 256.982 & 199.047 & 62.028\\
%2048 & 549.636 & 431.797 & 221.362 & 52.809\\\hline
2048 & 431.797 & 221.362 & 52.809\\\hline
\end{tabular}
\caption{Total memory utilization (in GBs) by the three largest data structures for the $512\lambda$ grid geometry.}
\label{fig:grid_memory_table}
\end{table*}

Table \ref{fig:grid_memory_table} shows the memory utilization of the three data structures with largest memory needs with increasing process counts. As expected, the memory used for tree storage (Tree Mem) does not increase with process count, despite some fluctuations due to different partitionings of the leaf nodes. This shows that the tree data structure is being nicely partitioned across processes. Size of the translation operators (Trans Ops) increase slowly with process count, slightly more than doubling going from 128 to 2048 processes. This is due to the spatial distribution of the tree nodes; multiple source observer pairs with the same translation in the tree may belong to different processes. Particularly, as the process count increases and the number of nodes in a process decreases. This results in some processes storing some of the same translation operators as the other processes. The greatest memory increase is in the message buffers (S/R Buffs). The translation send and receive buffers (S/R Buffs) are used to communicate the data for source nodes that interact with nodes in another process. Single node communications for each source and observer pair would eliminate the need for this buffer, but would result in drastically more translation messages which would degrade performance. So the translation message buffers are maximized to use any remaining memory to limit the number of translation messages that must be sent.

\begin{table*}
\centering
%\rowcolors{2}{gray!30}{white}
\begin{tabular}{|r|r|r|r|r|r|r|r|r|r|r|}
\hline
%$N_p$ & S/R Buffs & Trans Ops & Tree Mem & Tdir\\\hline
{\bf $N_p$} & {\bf S/R Buffs} & {\bf Trans Ops} & {\bf Tree Mem}\\\hline
%128 & 3.110 & 6.559 & 5.012 & 0.146\\
128 & 3.110 & 6.559 & 5.012\\
%256 & 8.402 & 10.095 & 5.161 & 0.292\\
256 & 8.402 & 10.095 & 5.161\\
%512 & 22.264 & 16.301 & 5.457 & 0.585\\
512 & 22.264 & 16.301 & 5.457\\
%1024 & 31.889 & 21.396 & 5.102 & 1.170\\
1024 & 31.889 & 21.396 & 5.102\\
%2048 & 51.736 & 30.390 & 5.414 & 2.340\\\hline
2048 & 51.736 & 30.390 & 5.414\\\hline
\end{tabular}
\caption{Total memory utilization (in GBs) by the three largest data structures for the $32\lambda$ volume geometry.}
\label{fig:volume_memory_table}
\end{table*}

On the other end of what can be performed with H-FMM is the volume geometry. Here the number of nodes per level is significantly increased due to the underlying full oct-tree structure (as opposed to a quad tree for a surface geometry), but the maximum height of a tree that can be computed is reduced. Most memory is reduced due to the shorter height of the tree, which reduces the size of the nodes at the top of the tree. However, the translation message buffers still use up as much memory as possible to improve translation communication performance.

\subsection{Performance Comparison with Other Codes}
Finally, we seek to compare our approach against those code that are available in the public domain. We note that open-source H-FMM codes are almost non-existent, with Abduljabbar et al. \cite{abduljabbar2019extreme} being a recent exception. Their BEMFMM code discretizes a mesh; the discretized points are used as particle inputs to our H-FMM code in order to compare processing of the same geometry. We ran a spherical geometry with 240 thousand mesh elements that produces 1.44 million points. Both codes are configured to produce a 6 level tree with slightly over 4100 leaf nodes with this sphere geometry and run with 64 nodes and 2048 MPI ranks. With this configuration, the BEMFMM implementation runs a single FMM iteration in 18.686 while our fine grain parallel FMM runs in 0.483 seconds. Our fine grain parallel Helmholtz FMM algorithm shows significantly faster performance in comparison.
%\metinnote{This part needs to be expanded a little bit: briefly describe BEMFMM, highlight its similarities to and differences from our code. For instance, it is a solver code but FMM time can be reported separately. They use a different meshing/discretization. How did you get the two codes working on roughly the same problems, what lambdas have you used for both? Are they single-precision, while we are using double? How much oversampling is done with each code -- so that we are getting similar accuracies?}
%\metinnote{what do we attribute this performance difference? Them doing local interpolation so having to use much more samples for similar accuracy? Them not parallelizing the higher level nodes very well? Other factors?}\mikenote{Added to this section, Steve's input would help as I'm rather fuzzy on some of the differences on how they are interpolating and similar.}

\section{Conclusions and Future Work}

We have demonstrated a novel method for parallel computation of large, upper level tree nodes in large-scale Helmholtz FMM which helps alleviate a key performance bottleneck associated with node dependency. The complexity of this method has been characterized. The results presented support the provided characterization and show the balance provided by this method.

Beyond the improvements presented, further work can be performed to improve memory usage. One possible method is a hybrid approach of MPI parallel with thread parallel to exploit shared memory parallelism to reduce duplicate memory allocation and communications.

\section*{Acknowledgment}
This research has in part been funded by the NSF Grant CCF-1822932. It has used computational resources at the National Energy Research Scientific Computing Center, a DOE Office of Science User Facility supported by the Office of Science of the U.S. Department of Energy under Contract No. DE-AC02-05CH11231, and at the High Performance Computing Center at Michigan State University. 

%\metinnote{I would expect to see a lot more references here - about 30 or more, given the extensive literature on Helmholtz FMM. Please try to populate it based on Steve's TAP paper or other relevant papers in your library.} 

%%%%%%%%%%%%%%%%%%%%%%%%%%%%%%%%%%%%%%%%%%%%%%%%
\bibliographystyle{ACM-Reference-Format}
\bibliography{bib/ParallelMLFMA,bib/bigbib2,bib/fmm,bib/bib_metin,bib/Shanker2016,bib/addRefs}


\begin{thebibliography}{00}

%%% the Web address.

\ifx \showCODEN    \undefined \def \showCODEN     #1{\unskip}     \fi
\ifx \showDOI      \undefined \def \showDOI       #1{#1}\fi
\ifx \showISBNx    \undefined \def \showISBNx     #1{\unskip}     \fi
\ifx \showISBNxiii \undefined \def \showISBNxiii  #1{\unskip}     \fi
\ifx \showISSN     \undefined \def \showISSN      #1{\unskip}     \fi
\ifx \showLCCN     \undefined \def \showLCCN      #1{\unskip}     \fi
\ifx \shownote     \undefined \def \shownote      #1{#1}          \fi
\ifx \showarticletitle \undefined \def \showarticletitle #1{#1}   \fi
\ifx \showURL      \undefined \def \showURL       {\relax}        \fi
% invisible to TeX
\providecommand\bibfield[2]{#2}
\providecommand\bibinfo[2]{#2}
\providecommand\natexlab[1]{#1}
\providecommand\showeprint[2][]{arXiv:#2}

\bibitem[\protect\citeauthoryear{Abduljabbar, Farhan, Al-Harthi, Chen, Yokota,
  Bagci, and Keyes}{Abduljabbar et~al\mbox{.}}{2019}]%
        {abduljabbar2019extreme}
\bibfield{author}{\bibinfo{person}{Mustafa Abduljabbar},
  \bibinfo{person}{Mohammed~Al Farhan}, \bibinfo{person}{Noha Al-Harthi},
  \bibinfo{person}{Rui Chen}, \bibinfo{person}{Rio Yokota},
  \bibinfo{person}{Hakan Bagci}, {and} \bibinfo{person}{David Keyes}.}
  \bibinfo{year}{2019}\natexlab{}.
\newblock \showarticletitle{Extreme scale FMM-accelerated boundary integral
  equation solver for wave scattering}.
\newblock \bibinfo{journal}{{\em SIAM Journal on Scientific Computing\/}}
  \bibinfo{volume}{41}, \bibinfo{number}{3} (\bibinfo{year}{2019}),
  \bibinfo{pages}{C245--C268}.
\newblock


\bibitem[\protect\citeauthoryear{Agullo, Bramas, Coulaud, Darve, Messner, and
  Takahashi}{Agullo et~al\mbox{.}}{2014}]%
        {agullo2014task}
\bibfield{author}{\bibinfo{person}{Emmanuel Agullo},
  \bibinfo{person}{B{\'e}renger Bramas}, \bibinfo{person}{Olivier Coulaud},
  \bibinfo{person}{Eric Darve}, \bibinfo{person}{Matthias Messner}, {and}
  \bibinfo{person}{Toru Takahashi}.} \bibinfo{year}{2014}\natexlab{}.
\newblock \showarticletitle{Task-based FMM for multicore architectures}.
\newblock \bibinfo{journal}{{\em SIAM Journal on Scientific Computing\/}}
  \bibinfo{volume}{36}, \bibinfo{number}{1} (\bibinfo{year}{2014}),
  \bibinfo{pages}{C66--C93}.
\newblock


\bibitem[\protect\citeauthoryear{Appel}{Appel}{1985}]%
        {Appel1985}
\bibfield{author}{\bibinfo{person}{A. Appel}.} \bibinfo{year}{1985}\natexlab{}.
\newblock \showarticletitle{An efficient program for many-body simulations}.
\newblock \bibinfo{journal}{{\em SIAM J. Sci. Comput.\/}}  \bibinfo{volume}{6}
  (\bibinfo{year}{1985}), \bibinfo{pages}{85--103}.
\newblock


\bibitem[\protect\citeauthoryear{Barnes and Hut}{Barnes and Hut}{1986}]%
        {Barnes1986}
\bibfield{author}{\bibinfo{person}{J. Barnes} {and} \bibinfo{person}{P. Hut}.}
  \bibinfo{year}{1986}\natexlab{}.
\newblock \showarticletitle{A hierarchical $((n \log n)$ force calculation
  algorithm}.
\newblock \bibinfo{journal}{{\em Nature\/}}  \bibinfo{volume}{324}
  (\bibinfo{year}{1986}), \bibinfo{pages}{446--449}.
\newblock


\bibitem[\protect\citeauthoryear{Cecka and Darve}{Cecka and Darve}{2013}]%
        {cecka2013fourier}
\bibfield{author}{\bibinfo{person}{Cris Cecka} {and} \bibinfo{person}{Eric
  Darve}.} \bibinfo{year}{2013}\natexlab{}.
\newblock \showarticletitle{Fourier-based fast multipole method for the
  Helmholtz equation}.
\newblock \bibinfo{journal}{{\em SIAM Journal on Scientific Computing\/}}
  \bibinfo{volume}{35}, \bibinfo{number}{1} (\bibinfo{year}{2013}),
  \bibinfo{pages}{A79--A103}.
\newblock


\bibitem[\protect\citeauthoryear{Chew, Michielssen, Song, and Jin}{Chew
  et~al\mbox{.}}{2001a}]%
        {Chew2001}
\bibfield{editor}{\bibinfo{person}{W.C. Chew}, \bibinfo{person}{E.
  Michielssen}, \bibinfo{person}{J.~M. Song}, {and} \bibinfo{person}{J.~M.
  Jin}} (Eds.). \bibinfo{year}{2001}\natexlab{a}.
\newblock \bibinfo{booktitle}{{\em Fast and Efficient Algorithms in
  Computational Electromagnetics}}.
\newblock \bibinfo{publisher}{Artech House, Inc.}, \bibinfo{address}{Norwood,
  MA, USA}.
\newblock
\showISBNx{1580531520}


\bibitem[\protect\citeauthoryear{Chew, Michielssen, Song, and Jin}{Chew
  et~al\mbox{.}}{2001b}]%
        {chew2001fast}
\bibfield{author}{\bibinfo{person}{Weng~Cho Chew}, \bibinfo{person}{Eric
  Michielssen}, \bibinfo{person}{JM Song}, {and} \bibinfo{person}{Jian-Ming
  Jin}.} \bibinfo{year}{2001}\natexlab{b}.
\newblock \bibinfo{booktitle}{{\em Fast and efficient algorithms in
  computational electromagnetics}}.
\newblock \bibinfo{publisher}{Artech House, Inc.}
\newblock


\bibitem[\protect\citeauthoryear{Dembart and Yip}{Dembart and Yip}{1995}]%
        {Dembart1995}
\bibfield{author}{\bibinfo{person}{B. Dembart} {and} \bibinfo{person}{E. Yip}.}
  \bibinfo{year}{1995}\natexlab{}.
\newblock \showarticletitle{A 3D fast multipole method for electromagnetics
  with multiple levels}. In \bibinfo{booktitle}{{\em Proceedings of the 11th
  Annual Conference on Applied Computational Electromagnetics}},
  Vol.~\bibinfo{volume}{1}. \bibinfo{address}{Monterey, CA},
  \bibinfo{pages}{621--628}.
\newblock


\bibitem[\protect\citeauthoryear{Dembart and Yip}{Dembart and Yip}{1998}]%
        {Dembart1998}
\bibfield{author}{\bibinfo{person}{B. Dembart} {and} \bibinfo{person}{E. Yip}.}
  \bibinfo{year}{1998}\natexlab{}.
\newblock \showarticletitle{The accuracy of fast multipole methods for
  Maxwell's equations}.
\newblock \bibinfo{journal}{{\em IEEE Computational Science and Engineering\/}}
   \bibinfo{volume}{5} (\bibinfo{year}{1998}), \bibinfo{pages}{48--56}.
\newblock


\bibitem[\protect\citeauthoryear{Ergul}{Ergul}{2011}]%
        {ergul2011parallel}
\bibfield{author}{\bibinfo{person}{Ozgur Ergul}.}
  \bibinfo{year}{2011}\natexlab{}.
\newblock \showarticletitle{{Parallel implementation of MLFMA for homogeneous
  objects with various material properties}}.
\newblock \bibinfo{journal}{{\em Progress In Electromagnetics Research\/}}
  \bibinfo{volume}{121} (\bibinfo{year}{2011}), \bibinfo{pages}{505--520}.
\newblock


\bibitem[\protect\citeauthoryear{Erg\"{u}l}{Erg\"{u}l}{2011}]%
        {ergul2011solutions}
\bibfield{author}{\bibinfo{person}{\"{O}zg\"{u}r Erg\"{u}l}.}
  \bibinfo{year}{2011}\natexlab{}.
\newblock \showarticletitle{{Solutions of large-scale electromagnetics problems
  involving dielectric objects with the parallel multilevel fast multipole
  algorithm}}.
\newblock \bibinfo{journal}{{\em JOSA A\/}} \bibinfo{volume}{28},
  \bibinfo{number}{11} (\bibinfo{year}{2011}), \bibinfo{pages}{2261--2268}.
\newblock


\bibitem[\protect\citeauthoryear{Erg{\"u}l and G{\"u}rel}{Erg{\"u}l and
  G{\"u}rel}{2008}]%
        {ergul2008hierarchical}
\bibfield{author}{\bibinfo{person}{{\"O} Erg{\"u}l} {and} \bibinfo{person}{L
  G{\"u}rel}.} \bibinfo{year}{2008}\natexlab{}.
\newblock \showarticletitle{Hierarchical parallelisation strategy for
  multilevel fast multipole algorithm in computational electromagnetics}.
\newblock \bibinfo{journal}{{\em Electronics Letters\/}} \bibinfo{volume}{44},
  \bibinfo{number}{1} (\bibinfo{year}{2008}), \bibinfo{pages}{3--5}.
\newblock


\bibitem[\protect\citeauthoryear{Ergul and Gurel}{Ergul and Gurel}{2013}]%
        {ergul2013accurate}
\bibfield{author}{\bibinfo{person}{Ozgur Ergul} {and} \bibinfo{person}{Levent
  Gurel}.} \bibinfo{year}{2013}\natexlab{}.
\newblock \showarticletitle{{Accurate solutions of extremely large
  integral-equation problems in computational electromagnetics}}.
\newblock \bibinfo{journal}{{\it Proc. IEEE}} \bibinfo{volume}{101},
  \bibinfo{number}{2} (\bibinfo{year}{2013}), \bibinfo{pages}{342--349}.
\newblock


\bibitem[\protect\citeauthoryear{Erg\"{u}l and G\"{u}rel}{Erg\"{u}l and
  G\"{u}rel}{2013}]%
        {ergul2013fast}
\bibfield{author}{\bibinfo{person}{\"{O}zg\"{u}r Erg\"{u}l} {and}
  \bibinfo{person}{Levent G\"{u}rel}.} \bibinfo{year}{2013}\natexlab{}.
\newblock \showarticletitle{{Fast and accurate analysis of large-scale
  composite structures with the parallel multilevel fast multipole algorithm}}.
\newblock \bibinfo{journal}{{\em JOSA A\/}} \bibinfo{volume}{30},
  \bibinfo{number}{3} (\bibinfo{year}{2013}), \bibinfo{pages}{509--517}.
\newblock


\bibitem[\protect\citeauthoryear{Greengard}{Greengard}{1988}]%
        {Greengard1988}
\bibfield{author}{\bibinfo{person}{L. Greengard}.}
  \bibinfo{year}{1988}\natexlab{}.
\newblock \bibinfo{booktitle}{{\em The rapid evaluation of potential fields in
  particle systems}}.
\newblock \bibinfo{publisher}{MIT Press}, \bibinfo{address}{Cambridge, MA}.
\newblock


\bibitem[\protect\citeauthoryear{Greengard, Huang, Rokhlin, and
  Wandzura}{Greengard et~al\mbox{.}}{1998}]%
        {Greengard1998}
\bibfield{author}{\bibinfo{person}{L. Greengard}, \bibinfo{person}{J. Huang},
  \bibinfo{person}{V. Rokhlin}, {and} \bibinfo{person}{S. Wandzura}.}
  \bibinfo{year}{1998}\natexlab{}.
\newblock \showarticletitle{Accelerating fast multipole methods for the
  Helmholtz equation at low frequencies}.
\newblock \bibinfo{journal}{{\em IEEE Computational Science and Engineering\/}}
   \bibinfo{volume}{5} (\bibinfo{year}{1998}), \bibinfo{pages}{32--38}.
\newblock


\bibitem[\protect\citeauthoryear{Greengard and Rokhlin}{Greengard and
  Rokhlin}{1987}]%
        {Greengard1987}
\bibfield{author}{\bibinfo{person}{L. Greengard} {and} \bibinfo{person}{V.
  Rokhlin}.} \bibinfo{year}{1987}\natexlab{}.
\newblock \showarticletitle{A fast algorithm for particle simulations}.
\newblock \bibinfo{journal}{{\it J. Comput. Phys.}}  \bibinfo{volume}{20}
  (\bibinfo{year}{1987}), \bibinfo{pages}{63--71}.
\newblock


\bibitem[\protect\citeauthoryear{Hamada, Narumi, Yokota, Yasuoka, Nitadori, and
  Taiji}{Hamada et~al\mbox{.}}{2009}]%
        {hamada2009gordonbell}
\bibfield{author}{\bibinfo{person}{Tsuyoshi Hamada}, \bibinfo{person}{Tetsu
  Narumi}, \bibinfo{person}{Rio Yokota}, \bibinfo{person}{Kenji Yasuoka},
  \bibinfo{person}{Keigo Nitadori}, {and} \bibinfo{person}{Makoto Taiji}.}
  \bibinfo{year}{2009}\natexlab{}.
\newblock \showarticletitle{42 TFlops Hierarchical N-body Simulations on GPUs
  with Applications in Both Astrophysics and Turbulence}. In
  \bibinfo{booktitle}{{\em Proceedings of the Conference on High Performance
  Computing Networking, Storage and Analysis}} {\em (\bibinfo{series}{SC
  '09})}. \bibinfo{publisher}{ACM}, \bibinfo{address}{New York, NY, USA},
  Article \bibinfo{articleno}{62}, \bibinfo{numpages}{12}~pages.
\newblock
\showISBNx{978-1-60558-744-8}
\showDOI{%
\url{https://doi.org/10.1145/1654059.1654123}}


\bibitem[\protect\citeauthoryear{Hughey, Aktulga, Melapudi, Shanker, Lu, and
  Michielssen}{Hughey et~al\mbox{.}}{2018}]%
        {hughey2018parallel}
\bibfield{author}{\bibinfo{person}{S Hughey}, \bibinfo{person}{HM Aktulga},
  \bibinfo{person}{V Melapudi}, \bibinfo{person}{B Shanker}, \bibinfo{person}{M
  Lu}, {and} \bibinfo{person}{E Michielssen}.} \bibinfo{year}{2018}\natexlab{}.
\newblock \showarticletitle{Parallel Non-Uniform MLFMA for Multiscale
  Electromagnetic Simulation}. In \bibinfo{booktitle}{{\em 2018 IEEE
  International Symposium on Antennas and Propagation \& USNC/URSI National
  Radio Science Meeting}}. IEEE, \bibinfo{pages}{1837--1838}.
\newblock


\bibitem[\protect\citeauthoryear{{Hughey}, {Aktulga}, {Vikram}, {Lu},
  {Shanker}, and {Michielssen}}{{Hughey} et~al\mbox{.}}{2019}]%
        {hughey2019parallel}
\bibfield{author}{\bibinfo{person}{S. {Hughey}}, \bibinfo{person}{H.~M.
  {Aktulga}}, \bibinfo{person}{M. {Vikram}}, \bibinfo{person}{M. {Lu}},
  \bibinfo{person}{B. {Shanker}}, {and} \bibinfo{person}{E. {Michielssen}}.}
  \bibinfo{year}{2019}\natexlab{}.
\newblock \showarticletitle{Parallel Wideband MLFMA for Analysis of
  Electrically Large, Nonuniform, Multiscale Structures}.
\newblock \bibinfo{journal}{{\em IEEE Transactions on Antennas and
  Propagation\/}} \bibinfo{volume}{67}, \bibinfo{number}{2}
  (\bibinfo{date}{Feb} \bibinfo{year}{2019}), \bibinfo{pages}{1094--1107}.
\newblock
\showISSN{1558-2221}
\showDOI{%
\url{https://doi.org/10.1109/TAP.2018.2882621}}


\bibitem[\protect\citeauthoryear{Ishiyama, Nitadori, and Makino}{Ishiyama
  et~al\mbox{.}}{2012}]%
        {ishiyama2012gordonbell}
\bibfield{author}{\bibinfo{person}{Tomoaki Ishiyama}, \bibinfo{person}{Keigo
  Nitadori}, {and} \bibinfo{person}{Junichiro Makino}.}
  \bibinfo{year}{2012}\natexlab{}.
\newblock \showarticletitle{4.45 Pflops Astrophysical N-body Simulation on K
  Computer: The Gravitational Trillion-body Problem}. In
  \bibinfo{booktitle}{{\em Proceedings of the International Conference on High
  Performance Computing, Networking, Storage and Analysis}} {\em
  (\bibinfo{series}{SC '12})}. \bibinfo{publisher}{IEEE Computer Society
  Press}, \bibinfo{address}{Los Alamitos, CA, USA}, Article
  \bibinfo{articleno}{5}, \bibinfo{numpages}{10}~pages.
\newblock
\showISBNx{978-1-4673-0804-5}
\showURL{%
\url{http://dl.acm.org/citation.cfm?id=2388996.2389003}}


\bibitem[\protect\citeauthoryear{Lingg, Hughey, and Aktulga}{Lingg
  et~al\mbox{.}}{2018}]%
        {lingg2018optimization}
\bibfield{author}{\bibinfo{person}{Michael~P Lingg}, \bibinfo{person}{Stephen~M
  Hughey}, {and} \bibinfo{person}{Hasan~Metin Aktulga}.}
  \bibinfo{year}{2018}\natexlab{}.
\newblock \showarticletitle{Optimization of the Spherical Harmonics Transform
  based Tree Traversals in the Helmholtz FMM Algorithm}. In
  \bibinfo{booktitle}{{\em Proceedings of the 47th International Conference on
  Parallel Processing}}. \bibinfo{pages}{1--11}.
\newblock


\bibitem[\protect\citeauthoryear{Melapudi, Shanker, Seal, and Aluru}{Melapudi
  et~al\mbox{.}}{2011}]%
        {melapudi2011scalable}
\bibfield{author}{\bibinfo{person}{Vikram Melapudi},
  \bibinfo{person}{Balasubramaniam Shanker}, \bibinfo{person}{Sudip Seal},
  {and} \bibinfo{person}{Srinivas Aluru}.} \bibinfo{year}{2011}\natexlab{}.
\newblock \showarticletitle{{A scalable parallel wideband MLFMA for efficient
  electromagnetic simulations on large scale clusters}}.
\newblock \bibinfo{journal}{{\em Antennas and Propagation, IEEE Transactions
  on\/}} \bibinfo{volume}{59}, \bibinfo{number}{7} (\bibinfo{year}{2011}),
  \bibinfo{pages}{2565--2577}.
\newblock


\bibitem[\protect\citeauthoryear{Michiels, Fostier, Bogaert, and {De
  Zutter}}{Michiels et~al\mbox{.}}{2013a}]%
        {michiels2013performing}
\bibfield{author}{\bibinfo{person}{Bart Michiels}, \bibinfo{person}{Jan
  Fostier}, \bibinfo{person}{Ignace Bogaert}, {and} \bibinfo{person}{Daniel {De
  Zutter}}.} \bibinfo{year}{2013}\natexlab{a}.
\newblock \showarticletitle{{Performing large full-wave simulations by means of
  a parallel MLFMA implementation}}. In \bibinfo{booktitle}{{\em Antennas and
  Propagation Society International Symposium (APSURSI), 2013 IEEE}}. IEEE,
  \bibinfo{pages}{1880--1881}.
\newblock


\bibitem[\protect\citeauthoryear{Michiels, Fostier, Bogaert, and {De
  Zutter}}{Michiels et~al\mbox{.}}{2013b}]%
        {michiels2013weak}
\bibfield{author}{\bibinfo{person}{Bart Michiels}, \bibinfo{person}{Jan
  Fostier}, \bibinfo{person}{Ignace Bogaert}, {and} \bibinfo{person}{Dani\"{e}l
  {De Zutter}}.} \bibinfo{year}{2013}\natexlab{b}.
\newblock \showarticletitle{{Weak scalability analysis of the
  distributed-memory parallel MLFMA}}.
\newblock \bibinfo{journal}{{\em Antennas and Propagation, IEEE Transactions
  on\/}} \bibinfo{volume}{61}, \bibinfo{number}{11} (\bibinfo{year}{2013}),
  \bibinfo{pages}{5567--5574}.
\newblock


\bibitem[\protect\citeauthoryear{Michiels, Fostier, Bogaert, and {De
  Zutter}}{Michiels et~al\mbox{.}}{2015}]%
        {michiels2015full}
\bibfield{author}{\bibinfo{person}{Bart Michiels}, \bibinfo{person}{Jan
  Fostier}, \bibinfo{person}{Ignace Bogaert}, {and} \bibinfo{person}{Daniel {De
  Zutter}}.} \bibinfo{year}{2015}\natexlab{}.
\newblock \showarticletitle{{Full-Wave Simulations of Electromagnetic
  Scattering Problems With Billions of Unknowns}}.
\newblock \bibinfo{journal}{{\em Antennas and Propagation, IEEE Transactions
  on\/}} \bibinfo{volume}{63}, \bibinfo{number}{2} (\bibinfo{year}{2015}),
  \bibinfo{pages}{796--799}.
\newblock


\bibitem[\protect\citeauthoryear{Michiels, Fostier, Bogaert, Demeester, and {De
  Zutter}}{Michiels et~al\mbox{.}}{2011}]%
        {michiels2011towards}
\bibfield{author}{\bibinfo{person}{Bart Michiels}, \bibinfo{person}{Jan
  Fostier}, \bibinfo{person}{Ignace Bogaert}, \bibinfo{person}{Piet Demeester},
  {and} \bibinfo{person}{Dani\"{e}l {De Zutter}}.}
  \bibinfo{year}{2011}\natexlab{}.
\newblock \showarticletitle{{Towards a scalable parallel MLFMA in three
  dimensions}}. In \bibinfo{booktitle}{{\em Computational Electromagnetics
  International Workshop (CEM), 2011}}. IEEE, \bibinfo{pages}{132--135}.
\newblock


\bibitem[\protect\citeauthoryear{Nishimura}{Nishimura}{2002}]%
        {nishimura2002fast}
\bibfield{author}{\bibinfo{person}{Naoshi Nishimura}.}
  \bibinfo{year}{2002}\natexlab{}.
\newblock \showarticletitle{Fast multipole accelerated boundary integral
  equation methods}.
\newblock \bibinfo{journal}{{\em Applied mechanics reviews\/}}
  \bibinfo{volume}{55}, \bibinfo{number}{4} (\bibinfo{year}{2002}),
  \bibinfo{pages}{299--324}.
\newblock


\bibitem[\protect\citeauthoryear{Rahimian, Lashuk, Veerapaneni,
  Chandramowlishwaran, Malhotra, Moon, Sampath, Shringarpure, Vetter, Vuduc,
  Zorin, and Biros}{Rahimian et~al\mbox{.}}{2010}]%
        {rahimian2010gordonbell}
\bibfield{author}{\bibinfo{person}{Abtin Rahimian}, \bibinfo{person}{Ilya
  Lashuk}, \bibinfo{person}{Shravan Veerapaneni}, \bibinfo{person}{Aparna
  Chandramowlishwaran}, \bibinfo{person}{Dhairya Malhotra},
  \bibinfo{person}{Logan Moon}, \bibinfo{person}{Rahul Sampath},
  \bibinfo{person}{Aashay Shringarpure}, \bibinfo{person}{Jeffrey Vetter},
  \bibinfo{person}{Richard Vuduc}, \bibinfo{person}{Denis Zorin}, {and}
  \bibinfo{person}{George Biros}.} \bibinfo{year}{2010}\natexlab{}.
\newblock \showarticletitle{Petascale Direct Numerical Simulation of Blood Flow
  on 200K Cores and Heterogeneous Architectures}. In \bibinfo{booktitle}{{\em
  Proceedings of the 2010 ACM/IEEE International Conference for High
  Performance Computing, Networking, Storage and Analysis}} {\em
  (\bibinfo{series}{SC '10})}. \bibinfo{publisher}{IEEE Computer Society},
  \bibinfo{address}{Washington, DC, USA}, \bibinfo{pages}{1--11}.
\newblock
\showISBNx{978-1-4244-7559-9}
\showDOI{%
\url{https://doi.org/10.1109/SC.2010.42}}


\bibitem[\protect\citeauthoryear{Salmon}{Salmon}{1991}]%
        {salmon1991parallel}
\bibfield{author}{\bibinfo{person}{John~K Salmon}.}
  \bibinfo{year}{1991}\natexlab{}.
\newblock {\em \bibinfo{title}{Parallel hierarchical N-body methods}}.
\newblock \bibinfo{thesistype}{Ph.D. Dissertation}. \bibinfo{school}{California
  Institute of Technology}.
\newblock


\bibitem[\protect\citeauthoryear{Sarvas}{Sarvas}{2003}]%
        {Sarvas2003}
\bibfield{author}{\bibinfo{person}{J. Sarvas}.}
  \bibinfo{year}{2003}\natexlab{}.
\newblock \showarticletitle{Performing Interpolation and Anterpolation by the
  Fast Fourier Transform in the 3D Multilevel Fast Multipole Algorithm}.
\newblock \bibinfo{journal}{{\em SIAM J. Numer. Anal.\/}}  \bibinfo{volume}{41}
  (\bibinfo{year}{2003}), \bibinfo{pages}{2180--2196}.
\newblock


\bibitem[\protect\citeauthoryear{Shanker and Huang}{Shanker and Huang}{2007}]%
        {Shanker2007}
\bibfield{author}{\bibinfo{person}{B. Shanker} {and} \bibinfo{person}{H.
  Huang}.} \bibinfo{year}{2007}\natexlab{}.
\newblock \showarticletitle{Accelerated Cartesian expansions - A fast method
  for computing of potentials of the form R\^{}\{- $\nu$\} for all real $\nu$}.
\newblock \bibinfo{journal}{{\it J. Comput. Phys.}}  \bibinfo{volume}{226}
  (\bibinfo{year}{2007}), \bibinfo{pages}{732--753}.
\newblock


\bibitem[\protect\citeauthoryear{Song, Lu, and Chew}{Song
  et~al\mbox{.}}{1997}]%
        {song1997}
\bibfield{author}{\bibinfo{person}{J.~M. Song}, \bibinfo{person}{C.~C. Lu},
  {and} \bibinfo{person}{W.~C. Chew}.} \bibinfo{year}{1997}\natexlab{}.
\newblock \showarticletitle{MLFMA for electromagnetic scattering by large
  complex objects}.
\newblock \bibinfo{journal}{{\em IEEE Transactions on Antennas and
  Propagation\/}}  \bibinfo{volume}{45} (\bibinfo{year}{1997}),
  \bibinfo{pages}{1488--1493}.
\newblock


\bibitem[\protect\citeauthoryear{Sundar, Sampath, and Biros}{Sundar
  et~al\mbox{.}}{2008}]%
        {sundar2008bottom}
\bibfield{author}{\bibinfo{person}{Hari Sundar}, \bibinfo{person}{Rahul~S
  Sampath}, {and} \bibinfo{person}{George Biros}.}
  \bibinfo{year}{2008}\natexlab{}.
\newblock \showarticletitle{Bottom-up construction and 2: 1 balance refinement
  of linear octrees in parallel}.
\newblock \bibinfo{journal}{{\em SIAM Journal on Scientific Computing\/}}
  \bibinfo{volume}{30}, \bibinfo{number}{5} (\bibinfo{year}{2008}),
  \bibinfo{pages}{2675--2708}.
\newblock


\bibitem[\protect\citeauthoryear{Taboada, Araujo, Basteiro, Rodr\'{\i}guez, and
  Landesa}{Taboada et~al\mbox{.}}{2013}]%
        {taboada2013mlfma}
\bibfield{author}{\bibinfo{person}{Jose~Manuel Taboada},
  \bibinfo{person}{Marta~G Araujo}, \bibinfo{person}{Fernando~Obelleiro
  Basteiro}, \bibinfo{person}{Jos\'{e}~Luis Rodr\'{\i}guez}, {and}
  \bibinfo{person}{Luis Landesa}.} \bibinfo{year}{2013}\natexlab{}.
\newblock \showarticletitle{{MLFMA-FFT parallel algorithm for the solution of
  extremely large problems in electromagnetics}}.
\newblock \bibinfo{journal}{{\it Proc. IEEE}} \bibinfo{volume}{101},
  \bibinfo{number}{2} (\bibinfo{year}{2013}), \bibinfo{pages}{350--363}.
\newblock


\bibitem[\protect\citeauthoryear{Velamparambil, Song, and Chew}{Velamparambil
  et~al\mbox{.}}{2000}]%
        {velamparambil2000parallelization}
\bibfield{author}{\bibinfo{person}{S Velamparambil}, \bibinfo{person}{Jiming
  Song}, {and} \bibinfo{person}{Weng~Cho Chew}.}
  \bibinfo{year}{2000}\natexlab{}.
\newblock \showarticletitle{On the parallelization of electrodynamic multilevel
  fast multipole method on distributed memory computers}. In
  \bibinfo{booktitle}{{\em Innovative Architecture for Future Generation
  High-Performance Processors and Systems, 1999. International Workshop}}.
  IEEE, \bibinfo{pages}{3--11}.
\newblock


\bibitem[\protect\citeauthoryear{Vikram, Knowles, Shanker, and Kempel}{Vikram
  et~al\mbox{.}}{2011}]%
        {Vikram2011}
\bibfield{author}{\bibinfo{person}{M. Vikram}, \bibinfo{person}{C. Knowles},
  \bibinfo{person}{B. Shanker}, {and} \bibinfo{person}{L.C. Kempel}.}
  \bibinfo{year}{2011}\natexlab{}.
\newblock \showarticletitle{An Ultra-Wideband FMM for Multi-scale
  Electromagnetic Simulations}.
\newblock \bibinfo{journal}{{\em 27th Annual Review of Progress in Applied
  Computational Electromagnetics\/}} (\bibinfo{year}{2011}).
\newblock


\bibitem[\protect\citeauthoryear{Vikram and Shanker}{Vikram and
  Shanker}{2009}]%
        {Vikram2009a}
\bibfield{author}{\bibinfo{person}{M. Vikram} {and} \bibinfo{person}{B.
  Shanker}.} \bibinfo{year}{2009}\natexlab{}.
\newblock \showarticletitle{An incomplete review of fast multipole methods from
  static to wideband as applied to problems in computational electromagnetics}.
\newblock \bibinfo{journal}{{\em Applied Computational Electromagnetics Society
  Journal\/}}  \bibinfo{volume}{27} (\bibinfo{year}{2009}),
  \bibinfo{pages}{79}.
\newblock


\bibitem[\protect\citeauthoryear{Waltz, Sertel, Carr, Usner, Volakis, and
  Others}{Waltz et~al\mbox{.}}{2007}]%
        {waltz2007massively}
\bibfield{author}{\bibinfo{person}{Caleb Waltz}, \bibinfo{person}{Kubilay
  Sertel}, \bibinfo{person}{Michael Carr}, \bibinfo{person}{Brian~C Usner},
  \bibinfo{person}{John~L Volakis}, {and} \bibinfo{person}{Others}.}
  \bibinfo{year}{2007}\natexlab{}.
\newblock \showarticletitle{{Massively parallel fast multipole method solutions
  of large electromagnetic scattering problems}}.
\newblock \bibinfo{journal}{{\em Antennas and Propagation, IEEE Transactions
  on\/}} \bibinfo{volume}{55}, \bibinfo{number}{6} (\bibinfo{year}{2007}),
  \bibinfo{pages}{1810--1816}.
\newblock


\bibitem[\protect\citeauthoryear{Wandzuraz}{Wandzuraz}{1993}]%
        {Wandzuraz1993}
\bibfield{author}{\bibinfo{person}{Stephen Wandzuraz}.}
  \bibinfo{year}{1993}\natexlab{}.
\newblock \showarticletitle{{The Fast Multipole Method for the Wave Equation: A
  Pedestrian Prescription}}.
\newblock \bibinfo{journal}{{\em IEEE Antennas and Propagation Magazine\/}}
  \bibinfo{volume}{35}, \bibinfo{number}{3} (\bibinfo{year}{1993}),
  \bibinfo{pages}{7--12}.
\newblock


\bibitem[\protect\citeauthoryear{Warren and Salmon}{Warren and Salmon}{1993}]%
        {Warren1993}
\bibfield{author}{\bibinfo{person}{M.S. Warren} {and} \bibinfo{person}{J.K.
  Salmon}.} \bibinfo{year}{1993}\natexlab{}.
\newblock \showarticletitle{A parallel hashed oct-tree N-body algorithm}. In
  \bibinfo{booktitle}{{\em Proc. Supercomputing}}. \bibinfo{pages}{1--12}.
\newblock


\bibitem[\protect\citeauthoryear{Yang, Wu, Gao, and Sheng}{Yang
  et~al\mbox{.}}{2019}]%
        {yang2019ternary}
\bibfield{author}{\bibinfo{person}{Ming-Lin Yang}, \bibinfo{person}{Bi-Yi Wu},
  \bibinfo{person}{Hong-Wei Gao}, {and} \bibinfo{person}{Xin-Qing Sheng}.}
  \bibinfo{year}{2019}\natexlab{}.
\newblock \showarticletitle{A Ternary Parallelization Approach of MLFMA for
  Solving Electromagnetic Scattering Problems with Over 10 Billion Unknowns}.
\newblock \bibinfo{journal}{{\em IEEE Transactions on Antennas and
  Propagation\/}} (\bibinfo{year}{2019}).
\newblock


\bibitem[\protect\citeauthoryear{Ying, Biros, and Zorin}{Ying
  et~al\mbox{.}}{2004}]%
        {ying2004kernel}
\bibfield{author}{\bibinfo{person}{Lexing Ying}, \bibinfo{person}{George
  Biros}, {and} \bibinfo{person}{Denis Zorin}.}
  \bibinfo{year}{2004}\natexlab{}.
\newblock \showarticletitle{A kernel-independent adaptive fast multipole
  algorithm in two and three dimensions}.
\newblock \bibinfo{journal}{{\it J. Comput. Phys.}} \bibinfo{volume}{196},
  \bibinfo{number}{2} (\bibinfo{year}{2004}), \bibinfo{pages}{591--626}.
\newblock


\bibitem[\protect\citeauthoryear{Ying and Zorin}{Ying and Zorin}{2004}]%
        {Ying2004}
\bibfield{author}{\bibinfo{person}{Lexing Ying} {and} \bibinfo{person}{Denis
  Zorin}.} \bibinfo{year}{2004}\natexlab{}.
\newblock \showarticletitle{A simple manifold-based construction of surfaces of
  arbitrary smoothness}.
\newblock \bibinfo{journal}{{\em ACM Transactions on Graphics\/}}
  \bibinfo{volume}{23} (\bibinfo{date}{August} \bibinfo{year}{2004}),
  \bibinfo{pages}{271--275}.
\newblock
Issue 3.
\showISSN{0730-0301}
\showDOI{%
\url{https://doi.org/10.1145/1015706.1015714}}


\end{thebibliography}

\end{document}